\documentclass[aps, prd, preprint, onecolumn, tightenlines, notitlepage, superscriptaddress, nofootinbib, preprintnumbers, floatfix,showkeys]{revtex4-1}

\usepackage{amstext}
\usepackage{amssymb}
\usepackage{amsmath}
\usepackage{rotating,graphicx}
\graphicspath{{plots/}}
\usepackage{hyperref}
\usepackage{url}
\usepackage{color}
\usepackage{ulem}
\usepackage[utf8]{inputenc}
\pdfoutput=1
\usepackage[x11names]{xcolor}
\usepackage{textcomp}

\usepackage{epsfig,amsfonts,mathrsfs,amsmath,amssymb,graphicx,color,slashed,multirow}
\usepackage{amsmath,latexsym,amssymb,graphicx,slashed,hyperref,color,enumerate,url,cancel,gensymb}
\usepackage{textcomp}

\hypersetup{colorlinks,citecolor= nicegreen,linkcolor= nicegreen}
\definecolor{nicered}{rgb}{0.7,0.1,0.1}
\definecolor{nicegreen}{rgb}{0.1,0.5,0.1}
\definecolor{mygreen}{rgb}{0,0.392,0}
\definecolor{mygreen}{rgb}{0,0,0.545}

\newcommand{\orcid}[1]{\href{https://orcid.org/#1}{#1}}

\AtBeginDocument{\hypersetup{citecolor=DarkOrange3,linkcolor=Green4,urlcolor=Green4}}
\begin{document}

\title{{\Large JUNO's  prospects for determining\\ the neutrino mass ordering }}

\author{David V. Forero}\email{dvanegas@udem.edu.co}
\thanks{orcid \# \orcid{0000-0003-4139-5670}}
\affiliation{Universidad de Medell\'{i}n, Carrera 87 N° 30 - 65 Medell\'{i}n, Colombia}
\author{Stephen J.~Parke}\email{parke@fnal.gov}  
\thanks{orcid \# \orcid{0000-0003-2028-6782}}
\affiliation{Theoretical Physics Dept., Fermi National Accelerator Laboratory, Batavia, IL, USA}
\author{Christoph~A.~Ternes}\email{ternes@to.infn.it}
\thanks{orcid \# \orcid{0000-0002-7190-1581}}
\affiliation{INFN, Sezione di Torino, Via P. Giuria 1, I--10125 Torino, Italy}
\author{Renata Zukanovich Funchal}\email{zukanov@fma.if.usp.br} 
\thanks{orcid \# \orcid{0000-0001-6749-0022}}
\affiliation{Instituto de F\'{i}sica, Universidade de S\~{a}o  Paulo, 
S\~{a}o Paulo, Brazil\\[5mm] ~}

\preprint{FERMILAB-PUB-21-201-T}

\begin{abstract}

The flagship measurement of the JUNO experiment is the determination of the neutrino mass ordering.  Here we revisit  its prospects to make this determination by 2030, using the current global knowledge of the relevant neutrino parameters as well as current information on the reactor configuration and the critical parameters of the JUNO detector. 
We pay particular attention to the non-linear detector energy response. 
Using the measurement of $\theta_{13}$ from Daya Bay,  but without information from other experiments, we estimate  the probability of JUNO determining the 
neutrino mass ordering at $\ge$  3$\sigma$ to be  31\% by 2030.
As this probability is particularly sensitive to the true values of the oscillation parameters, especially  $\Delta m^2_{21}$,
JUNO's improved measurements of $\sin^2 \theta_{12}$, $\Delta m^2_{21}$ and $|\Delta m^2_{ee}|$,   obtained after a couple of years of operation,
 will allow an updated estimate of the probability that JUNO alone can determine the neutrino mass ordering  by the end of the decade.
Combining JUNO's measurement of $|\Delta m^2_{ee}|$ with other experiments in a global fit will  most likely lead to an earlier determination of the mass ordering.

\end{abstract}

\keywords{Neutrino Physics, JUNO.}
\maketitle
\newpage
\tableofcontents

\section{Introduction}
\label{sec:intro}
After the first observation of the so-called {\it solar neutrino puzzle} by the Homestake experiment in the late 60's, it took us about 30 years to establish that neutrino flavor oscillations are prevalent in nature, impacting cosmology, astrophysics as well as nuclear and particle physics. 
In the last 20 years we have consolidated our understanding of neutrino oscillations both at the experimental as well as the theoretical level. 
We know now, thanks to a great number of experimental efforts involving solar, atmospheric, accelerator and reactor neutrino oscillation experiments, that neutrino oscillations are genuine three flavor phenomena driven by two independent mass squared differences ($\Delta m^2_{21}$ and $\Delta m^2_{32}$) and three mixing angles ($\theta_{12}$, $\theta_{13}$ and $\theta_{23}$) and possibly a charge-parity violating phase ($\delta_{\rm CP}$). See \cite{Nunokawa:2007qh} for a review of how these parameters are defined.

Today a single class of experiments dominates the precision of the measurement of each of these aforementioned parameters~\cite{deSalas:2020pgw,Capozzi:2021fjo,Esteban:2020cvm}. 
In the solar sector ($12$), $\Delta m^2_{21}$ is determined to less than 3\% mainly by the 
KamLAND~\cite{Gando:2013nba} long-baseline $\bar \nu_e$  disappearance reactor experiment, while $\sin^2\theta_{12}$ is determined by the combination of solar neutrino experiments~\cite{Cleveland:1998nv,Kaether:2010ag,Abdurashitov:2009tn,Bellini:2011rx,Bellini:2013lnn,Hosaka:2005um,Cravens:2008aa,Abe:2010hy,Nakano:PhD,yasuhiro_nakajima_2020_4134680,Aharmim:2011vm,Ahmad:2002jz} to $\sim 4\%$. 
In the atmospheric sector ($23$), $\Delta m^2_{32}$ (or $\Delta m^2_{31}$) and $\sin^2\theta_{23}$ are dominantly determined  by the $\nu_\mu$ and $\bar \nu_\mu$ disappearance  accelerator experiments MINOS~\cite{Adamson:2013ue}, NOvA~\cite{Acero:2019ksn,alex_himmel_2020_3959581} and T2K~\cite{Abe:2021gky}, with corresponding precision of better than 1.5\% and 8\%, respectively. 
The mixing $\sin^2\theta_{13}$, that connects the solar and atmospheric sectors, is determined by 
the short-baseline $\bar \nu_e$ disappearance reactor experiments Daya Bay~\cite{Adey:2018zwh}, RENO~\cite{Bak:2018ydk,jonghee_yoo_2020_4123573} and Double Chooz~\cite{DoubleChooz:2019qbj} to a precision of $\sim 3\%$. 
Regarding the CP-phase $\delta_{\rm CP}$ there is a small tension in the determination among current experiments T2K and NOvA~\cite{deSalas:2020pgw,Esteban:2020cvm,Kelly:2020fkv}. 
The determination of $\delta_{\rm CP}$ remains an open problem that probably will have to be addressed by the next generation of long-baseline neutrino experiments such as DUNE and Hyper-K.
There is, nevertheless, an important open question that  influences the better determination of some of these parameters: what is the neutrino mass ordering?

If one defines the mass eigenstates $\nu_1$, $\nu_2$, $\nu_3$ in terms of decreasing  amount of electron neutrino flavor content,  then the results of the SNO solar neutrino experiment determined that  $m_1<m_2$. 
However, the available information does not allow us to know the complete ordering yet: both $m_1<m_2<m_3$ (normal ordering, NO) and $m_3<m_1<m_2$ (inverted ordering, IO) are compatible with the current data~\cite{deSalas:2020pgw,Esteban:2020cvm,Kelly:2020fkv}.
The measurement of the neutrino mass ordering is one of the most pressing and delicate challenges of our times.
Besides its direct impact on the precise knowledge of the oscillation parameters, neutrino mass ordering affects the sum of neutrino masses from cosmology, the search for neutrinoless double-$\beta$ decay and ultimately, our better understanding of the pattern of masses and mixing in the leptonic sector.

The use of $\bar \nu_e$ from nuclear reactors with a medium-baseline detector to determine the mass ordering, exploring genuine three generation effects as long as $\sin^2\theta_{13} \gtrsim$ few \%,  was first proposed in~\cite{Petcov:2001sy}. 
This idea was further investigated in~\cite{Choubey:2003qx} for a general experiment and more recently in~\cite{Bilenky:2017rzu}, specifically  for JUNO.
In all three of these papers, different artificial constraints were imposed on the $\Delta m^2_{3i}$'s when comparing the NO and IO spectra.  As we will see, any and all of these artificial constrains increases the difference between the NO and IO, see appendix \ref{appx:artificial} for a more detailed discussion.  
In fact, it was shown in Ref.~\cite{Minakata:2007tn} that what these experiments can precisely measure is the effective combination~\cite{Nunokawa:2005nx} 
\begin{equation}
\Delta m^2_{ee} \equiv \cos^2 \theta_{12} \Delta m^2_{31}+ \sin^2\theta_{12} \Delta m^2_{32}\,,
\label{eq:dmsqee}
\end{equation}
and the sign ($+$ for NO, $-$ for IO) of a phase ($\Phi_\odot$) that depends on the solar parameters.
This subtlety is of crucial importance in correctly assessing the sensitivity to the neutrino mass ordering.

The Jiangmen Underground Neutrino Observatory (JUNO)~\cite{An:2015jdp}, a 20~kton liquid scintillator detector located in the Guangdong Province at about 53~km from the Yangjiang and Taishan nuclear power plants in China, will be the first experiment to implement this idea.
This medium-baseline facility offers the unprecedented opportunity to access in a single experiment four of the oscillation parameters: $\Delta m^2_{21}$, $\sin^2\theta_{12}$, $\vert \Delta m^2_{ee}\vert$, $\sin^2\theta_{13}$ and the sign of phase advance, $\Phi_\odot(L/E)$, which determines the mass ordering.  
JUNO aims in the first few years to measure $\Delta m^2_{21}$, $\sin^2\theta_{12}$ and $\vert \Delta m^2_{ee} \vert$ with a precision $\lesssim 1\%$ to be finally able, after approximately 8 years\footnote{ Assuming 26.6 GW of reactor power. The original 6 years assumed 35.8 GW.},  to  determine the neutrino mass ordering at 3$\sigma$ confidence level (C.L.).

Many authors have studied the neutrino mass ordering determination at medium-baseline reactor neutrino experiments such as JUNO. 
In Ref.~\cite{Zhan:2008id} a Fourier analysis was proposed, but no systematic effects were considered.
The effect of energy resolution was investigated in Ref.~\cite{Ge:2012wj}.
The importance of also taking into account non-linear effects in the energy spectrum reconstruction was first pointed out in Ref.~\cite{Parke:2008cz} and 
addressed in Ref.~\cite{Qian:2012xh,Li:2013zyd}, where limited impact on the mass ordering was observed.
Matter effects, geo-neutrino background, energy resolution, energy-scale and spectral shape uncertainties were investigated in \cite{Capozzi:2013psa}.
The impact of the energy-scale and flux-shape uncertainties was further explored in \cite{Capozzi:2015bpa}. 
The benefits of a near detector for JUNO was demonstrated in \cite{Forero:2017vrg} and in \cite{Cheng:2020ivh} the impact of the sub-structures in the reactor antineutrino spectrum, due to Coulomb effects in beta decay, was studied in the light of a near detector under various assumptions of the detector energy resolution.  This was further explored in~\cite{Capozzi:2020cxm}.
In \cite{Blennow:2013oma} the distribution for the test statistics, to address the mass ordering determination, was proven to be normally distributed and this was also applied to quantify the JUNO sensitivity. It was also shown that without statistical fluctuations, the mentioned test statistics is equivalent to the widely adopted $\Delta \chi^2$ approach used in sensitivity studies.
Finally, the combined sensitivity of JUNO and PINGU was also recently studied by the authors of \cite{Bezerra:2019dao}, while a combined sensitivity study of JUNO and T2K or NOvA was performed in~\cite{Cabrera:2020own}.

One can appreciate the difficulty in establishing the mass ordering with this setup by noticing that after 8 years (2400 days) of data taking, the difference in the number of events for NO and IO  is only a few tens of events per energy bin which is smaller than the statistical uncertainty in each bin. 
It is clear that this formidable endeavor depends on stringent requirements on the experiment's systematic uncertainties, but also on the actual values of the oscillation parameters as well as on statistical fluctuations. 
This is why we think it is meaningful to revisit the prospect that JUNO can obtain a 3$\sigma$ preference of the neutrino mass ordering by 2030.
This is the task we undertake in this paper.

Our paper is structured as follows. In Sec.~\ref{sec:osc} we describe the $\bar \nu_e$ survival probability in a way that highlights the physics that is relevant for medium-baseline reactor neutrino experiments and how it depends on the oscillation parameters. 
In Sec.~\ref{sec:sim} we explain how we simulate the experiment and show the statistical challenges associated with extracting the mass ordering in a medium-baseline reactor experiment.
Sec.~\ref{sec:massordering} addresses how the following experimental details affect the determination power of the neutrino mass ordering of the JUNO experiment;
(A) reactor distribution and backgrounds, (B) bin to bin flux uncertainties, (C) the number of energy bins used in the analysis, (D) the size of the energy resolution.
In Sec.~\ref{sec:osc_effect} we show how varying the true values of the neutrino oscillation parameters improves or reduces the prospects for JUNO's determination of the neutrino mass ordering.
Sec.~\ref{sec:NLeffects} addresses the effects of the non-linear detector response on the mass ordering determination.
In Sec.~\ref{sec:fluct} we simulate 60\,k experiments consistent with the current best fit values and uncertainties of the oscillation parameters. From this simulation we estimate the probabilities that JUNO can determine the mass ordering at $\ge 3\sigma$ with 4, 8 and 16 years of data taking. 
In Sec.~\ref{sec:combined} we show how JUNO's measurement of $\Delta m^2_{ee}$, when combined with other experiments can determine the mass ordering after a few years of data taking.
Finally in Sec.~\ref{sec:conc} we draw our conclusions.
There are four Appendices: Appendix \ref{appx:artificial} addresses the effects of imposing artificial constraints on the $\Delta m^2$'s,  Appendix \ref{appx:Prob}
gives a derivation of the oscillation probability used in this paper, Appendix \ref{app:compare} compares our  analysis with the JUNO collaboration's  analysis 
and Appendix  \ref{appx:contrib} discusses the current impact of T2K, NOvA and the  atmospheric neutrino data on the determination of $\vert \Delta m^2_{ee}\vert$.

\section{The $\bar\nu_e$   survival probability}
\label{sec:osc}

The neutrino survival probability for reactor experiments in vacuum is given by
\begin{equation}
 P_{\overline{\nu}_e\to\overline{\nu}_e} = 1 - \sin^22\theta_{13}\left[ \cos^2 \theta_{12} \sin^2\Delta_{31}+  \sin^2 \theta_{12} \sin^2\Delta_{32}\right]-P_\odot\,,
 \label{eq:prob0}
\end{equation}
where the  kinematic phases are  $\Delta_{ij} \equiv \Delta m_{ij}^2L/(4E)$ and  
$P_\odot = \sin^22\theta_{12}\cos^4\theta_{13}\sin^2\Delta_{21}$.
This survival probability was first rewritten, without approximation, in a more useful way for the medium baseline reactor experiments in \cite{Minakata:2007tn}, as
\begin{equation}
 P_{\overline{\nu}_e\to\overline{\nu}_e} = 1 - \frac{1}{2}\sin^22\theta_{13}\left[1-\sqrt{1-\sin^22\theta_{12}\sin^2\Delta_{21}}\, \cos(2|\Delta_{ee}|\pm \Phi_\odot)\right]-P_\odot\,,
 \label{eq:prob}
\end{equation}
where $\Delta m^2_{ee}$, defined in Eq.~\eqref{eq:dmsqee},  is the effective atmospheric $\Delta m^2$ for $\nu_e$ disappearance, see \cite{Nunokawa:2005nx,Parke:2016joa}.  
The mass ordering is determined by the sign in front of $\Phi_\odot$, `+' (`--') for NO (IO).
The phase advance or retardation $\Phi_\odot$ is
\begin{align}
 \Phi_\odot  =& \arctan\left(\cos2\theta_{12}\tan\Delta_{21}\right) - \Delta_{21}\cos2\theta_{12}\, .  
 \label{eq:phiodot}
 \end{align}
Note that the survival probability depends only on four of the oscillation parameters, $\theta_{12}$, $\theta_{13}$, $\Delta m_{21}^2$ and $\vert \Delta m_{ee}^2\vert$ and the sign for the mass ordering. The determination of the sign of $ \Delta m_{21}^2 ~\cos 2\theta_{12} > 0 $  by SNO~\cite{Aharmim:2005gt} is crucial for this measurement. See Appendix \ref{appx:Prob} for more details on the survival probability.

\begin{figure}[t]
  \centering
  \includegraphics[width=0.65\textwidth]{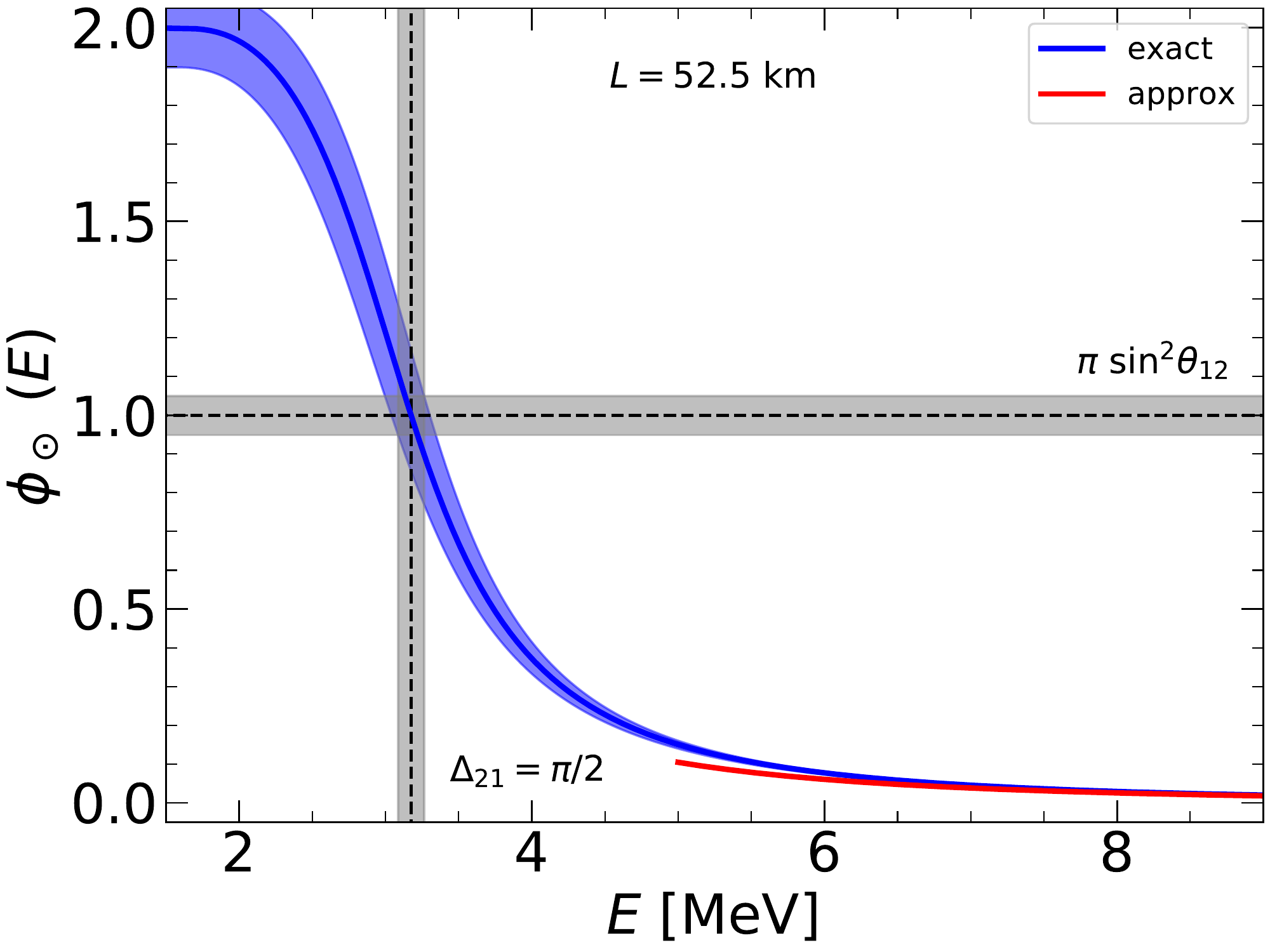}
    \caption{ The kinematic phase advance/retardation,  $\Phi_\odot$,  of the survival probability as a function of $E$  at $L=52.5$~km. The blue band is obtained from the exact formula, while the red curve shows the approximation for values of $L/E< 10$\,km/MeV. The dashed vertical and horizontal lines mark the solar oscillation minimum, i.e. $\Delta _{21} = \pi/2$, where $\Phi_\odot = \pi\,\sin^2\theta_{12}\approx 0.999$. The gray and blue bands  are obtained by varying the solar parameters in their corresponding 1$\sigma$ intervals as given in Table \ref{tab:oscparam}. $\Phi_\odot$ as function of $L/E$ is given in Appendix~\ref{appx:Prob}.}
  \label{fig:phi_fig}
\end{figure}

For $\Delta_{21} \ll \pi/2$, the phase advance/retardation can be approximated by 

\begin{equation}
\Phi_\odot \approx \frac1{3} \, \sin^2 2 \theta_{12} \, \cos 2 \theta_{12} \, \Delta^3_{21} +{\cal O}(\Delta^5_{21})\, , 
\end{equation}
and then near $\Delta_{21}\approx 1$ rises rapidly so that

\begin{equation}
\Phi_\odot(\Delta_{21}=\pi/2) = \pi  \sin^2 \theta_{12} \approx 1\,.  
\end{equation}
This behavior is illustrated in Fig.~\ref{fig:phi_fig}, using the central and 1$\sigma$ bands for the solar parameters given in Table \ref{tab:oscparam} taken from the recent global fit  \cite{deSalas:2020pgw}. 
Here, we show the advance/retardation as a function of $E$ at $L=52.5$~km, and in Appendix \ref{appx:Prob} also as function of $L/E$.

\begin{table}[b]
\centering
  \catcode`?=\active \def?{\hphantom{0}}
   \begin{tabular}{|c|c|c|}
   \hline
   \multicolumn{3}{|c|}{Normal Ordering}\\
    \hline\hline
    Parameter & Nominal Value & $1 \sigma$
    \\
    \hline
    $\sin^2\theta_{12}$     & 0.318 & $\pm0.016$\\ 
    $\Delta m^2_{21}$[$10^{-5}$eV$^2$]       & $7.50$ & $  \pm0.21$\\[1mm]  
    \hline  
        $\sin^2\theta_{13}$     & 0.02200 & $\pm0.00065$ \\
    $\Delta m^2_{ee}$ [$10^{-3}$eV$^2$]        &  $2.53$ & $+0.03$/$-0.02$\\
    \hline
    \end{tabular}
    \caption{Nominal values and uncertainties of the neutrino oscillation parameters used to simulate data in this paper. Throughout the paper we simulate data assuming NO. 
    These values were taken from the global fit to neutrino oscillation data found in Ref.~\cite{deSalas:2020pgw}.}
    \label{tab:oscparam} 
\end{table}

Matter effects are important in the determination of the best fit values of the solar parameters $\Delta m^2_{21}$ and $\sin^2 \theta_{12}$. 
The size of these effects, which do not satisfy the naive expectations, was first given in a numerical simulation in~\cite{Li:2016txk}, and later explained in a semi-analytical way in~\cite{Khan:2019doq}.  
For $\Delta m^2_{21}$ and $\sin^2 \theta_{12}$, the sizes of these shifts are -1.1\% and 0.20\%, respectively. 
Since we are interested only in sensitivities, we can ignore matter effects in the propagation of neutrinos in this paper and will use here the vacuum expression for the survival probability, Eq.~\eqref{eq:prob}. 
However, in a full analysis of real data matter effects must be included.
In the next section we will describe details of our simulation of the JUNO reactor experiment.

\section{Simulation of a medium baseline reactor experiment}
\label{sec:sim}
 For the simulation of JUNO we use the information given in Refs.~\cite{An:2015jdp,Bezerra:2019dao} but with the updates of baselines, efficiencies and backgrounds provided in Ref.~\cite{Abusleme:2021zrw}.
In order to simulate the event numbers and to perform the statistical analysis we use the GLoBES software~\cite{Huber:2004ka,Huber:2007ji}. 
We start with an idealized configuration where all reactors that provide the 26.6 GW$_{\text{th}}$ total thermal power are at 52.5~km baseline from the detector.
The antineutrinos are mainly created in the fission of four isotopes,  $^{235}$U (56.1\%), $^{238}$U (7.6\%), $^{239}$Pu (30.7\%) and $^{241}$Pu (5.6\%)~\cite{Abusleme:2020bzt}.
For our simulation we use the Huber-Mueller flux predictions~\cite{Mueller:2011nm, Huber:2011wv} for each isotope. 
The $\bar \nu_e$ propagate to the JUNO detector and are observed via inverse beta decay $\overline{\nu}_e + p \rightarrow e^+ + n$~\cite{Vogel:1999zy}.
We assume a liquid scintillator detector with a 20~kton fiducial mass and a running time of 2400 days (8 years @ 82\%  live time)\footnote{An exposure of 26.6 GW$_\text{th}$ for 2400 days (8 years @ 82\%) is equivalent to  35.8 GW$_\text{th}$ for 1800 days (6 years @ 82\%) as used in \cite{An:2015jdp}.}.
We will include the detector energy resolution of 3.0\% unless otherwise stated. 
The aforementioned quantities affect the calculation of  the event numbers. The number of events, $N_i$, in the $i$-th bin corresponding to the reconstructed neutrino energy $E_i$ is given by

\begin{equation}
  N_i = \mathcal{N_T}\int dE \int_{E_i^{\text{min}}}^{E_i^{\text{max}}} dE'~\phi_{\overline{\nu}_e}(E)~P_{\overline{\nu}_e\to\overline{\nu}_e}(E,L)~\sigma(E)~R(E,E_i')\,.
\end{equation}
Here, $\mathcal{N_T}$ is a normalization constant taking into account the exposure time, efficiency, fiducial mass of the detector and reactor-detector distance, $\phi_{\overline{\nu}_e}(E)$ is the antineutrino flux, $ P_{\overline{\nu}_e\to\overline{\nu}_e}(E,L)$ is the survival probability in Eq.~\eqref{eq:prob}, $\sigma(E)$ is the cross section, and $R(E,E_i')$ is the energy resolution function 

\begin{equation}
 R(E,E') = \frac{1}{\sqrt{2\pi}\sigma_E(E)}\exp\left(-\frac{(E-E')^2}{2\sigma_E^2(E)}\right)\,,
\end{equation}
which relates the reconstructed and true neutrino energies. The energy resolution is given by

\begin{equation}
  \sigma_E(E) = \epsilon ~ \sqrt{ E_p/\rm MeV}~\text{MeV} \,, 
  \end{equation}
where the prompt energy $E_p$ is given by 
$$E_p= E- \Delta M, \quad {\rm with } \quad \Delta M \equiv m_n-m_p-m_e= 0.78~\text{MeV}.$$
The variable $\epsilon$ is  the detector energy resolution. In this paper we will use 
$\epsilon = 3.0\% $ except when  discussing the effects of varying this parameter in Sec.~\ref{sec:massordering} D, where
we also  will use 2.9\% and 3.1\%.

\begin{figure}[t]
  \centering
  \includegraphics[width=0.53\textwidth]{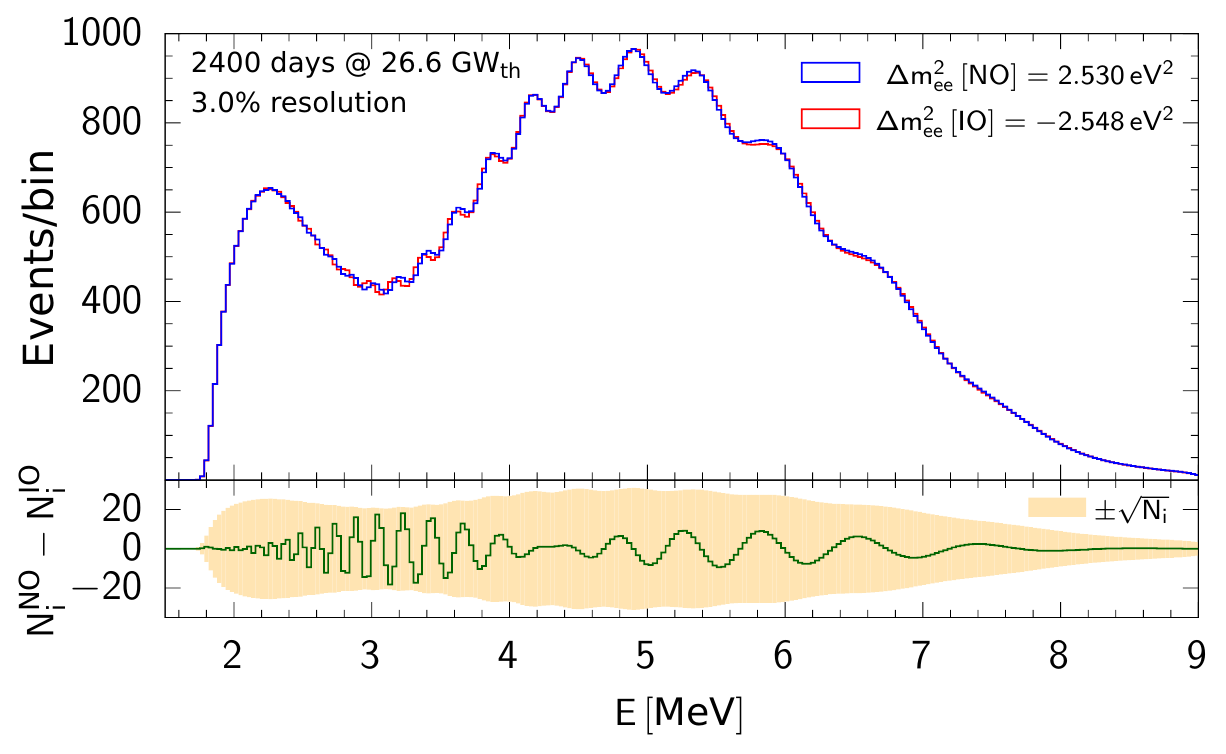}    \quad
      \includegraphics[width=0.43\textwidth]{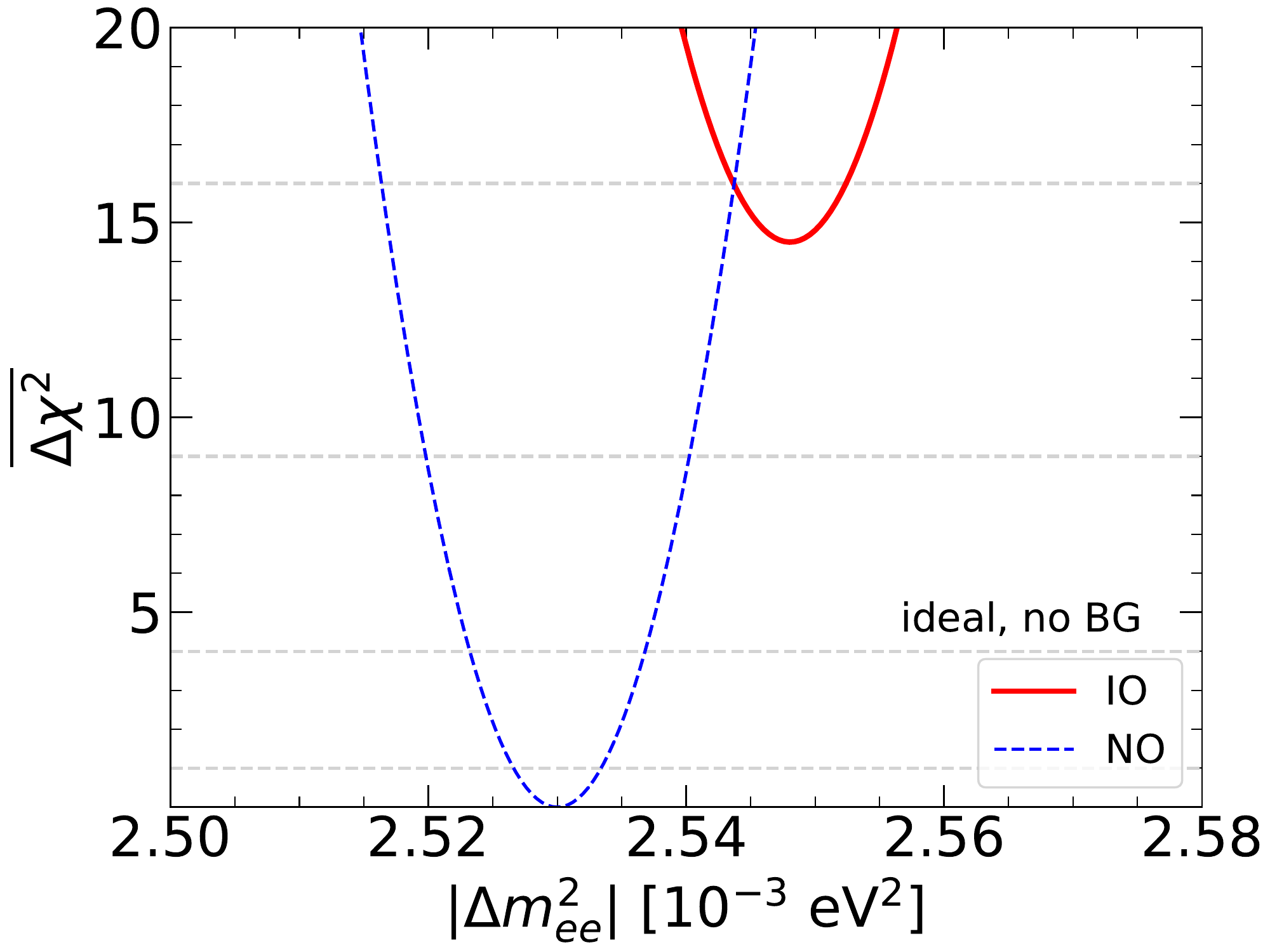}
      \caption{
  In the upper left panel we show the oscillated spectra for NO (blue) and for IO (red) for 8 years (2,400 live days) of data using 26.6 GW$_{\text{th}}$ with all core-detector baselines set at 52.5~km.  No systematic effects and no backgrounds are included. There are 200 bins between 1.8 and 8.0 MeV, with a bin size of 31 keV, and 3.0\% resolution was used. While $\Delta m^2_{ee}~[\text{NO}]$ is the input, $\Delta m^2_{ee}~[\text{IO}]$ is chosen to minimize the statistical $ \overline{\chi^2}$ between the two spectra, see right panel  ($ \overline{\chi^2}_{\rm min}[\text{IO}] = 14.5$, see right panel). 
  The parameters $\sin^2 \theta_{13}$, $\sin^2 \theta_{12}$ and $\Delta m^2_{21}$ are from Table~\ref{tab:oscparam}. In the left lower panel, the difference between the two oscillated spectra in each bin (green),  $ N^{\rm NO}_i - N^{\rm IO}_i$, is given, as well as plus/minus  statistical uncertainty in each  oscillated bin (orange band), $  \pm\sqrt{N^{\rm NO}_i} \approx \pm \sqrt{ N^{\rm IO}_i}$. Note, the difference is always within the statistical uncertainty for that bin. }
  \label{fig:spectra}
\end{figure}

In Fig.~\ref{fig:spectra} we have plotted the event spectrum for JUNO using 200 bins for 8 years  (2,400 live days) of data taking  and 26.6\,GW$_\text{th}$. 
In the top panel, the blue and red spectra corresponds to  
\begin{equation*}
 \Delta m_{ee}^2~~[\text{NO}] = 2.530\times 10^{-3} ~\text{eV}^2 ~~\text{and} ~~\Delta m_{ee}^2~~[\text{IO}] = -2.548\times 10^{-3}~ \text{eV}^2\,,
\end{equation*}
respectively\footnote{Note, that the value for $\Delta m_{ee}^2~[\text{IO}]$ does not correspond to any of the artificial constraints on the atmospheric mass splitting imposed in Refs.~\cite{Petcov:2001sy,Choubey:2003qx,Bilenky:2017rzu}, see Appendix~\ref{appx:artificial} for more details.}. 
The $\Delta m_{ee}^2$ for NO is input whereas the value for IO is chosen so as to minimize the 
 $\overline{\Delta \chi^2}=\overline{\chi^2}_\text{min}[\rm IO]-\overline{\chi^2}_\text{min}[\rm NO]$ between the two spectra. 
 By construction $\overline{\chi^2}_\text{min}[\rm NO]=0$, so minimizing $\overline{\Delta \chi^2}$ is equivalent to minimizing $\overline{\chi^2}_\text{min}[\rm IO]$.
In the lower panel, we plot the difference in event spectra obtained for NO and IO. Note that this difference is less than 20 events/bin. 
Also shown is the statistical uncertainty in each oscillated bin (orange band), which for all bins exceeds the difference between the NO and IO event 
spectra\footnote{Caveat: if one halves the bin size in this figure the difference between the NO and IO goes down by a factor of 2 whereas the statistical uncertainty by only $\sqrt{2}$ making the difference more challenging to observe. 
If one doubles the bin size the statistical uncertainty increases by the $\sqrt{2}$ whereas the difference would increase by 2, this improves the situation except for the fact that at low energy there is some washing out of the difference.}.
This figure demonstrates the statistical challenges for JUNO to determine the mass ordering and will be addressed in more detail in Sec. \ref{sec:fluct}. 
For reference, we also show on the right panel  of Fig.~\ref{fig:spectra} the corresponding $\overline{\chi^2}$ distributions. Throughout this paper we will use 
dashed (solid) lines for the fit with NO (IO).

Note that including systematic uncertainties as well as the 
real distribution of core-reactor distances  and backgrounds will further decrease the difference between the two spectra.
But first let us address the simulation details and systematic uncertainties. 

To perform the statistical analysis we create a spectrum of fake data $N_{i}^\text{dat}$ for some set of oscillation parameters.
Next we try to reconstruct this spectrum varying the relevant oscillation parameters $\vec{p}$. 
For each set $\vec{p}$ we calculate a $\chi^2$ function

\begin{equation}
 \chi^2(\vec{p}) =\min_{\vec{\alpha}}\sum_{i} \frac{(N_{i}^\text{dat} - N_{i}(\vec{p},\vec{\alpha}))^2}{N_{i}^\text{dat}} + \sum_j\left(\frac{\alpha_j}{\sigma_j}\right)^2 + \chi^2_{\text{NL}},
 \label{eq:chi2}
\end{equation}
where $N_i(\vec{p},\vec{\alpha})$ is the predicted number of events\footnote{The number of events includes the background events extracted from Ref.~\cite{An:2015jdp}.} for parameters $\vec{p}$, $\vec{\alpha} = (\alpha_1,\alpha_2,\ldots)$ are the systematic uncertainties with their corresponding standard deviations $\sigma_k$. $\chi^2_{\text{NL}}$ is the penalty for the non-linear detector response and will be discussed in more detail in Sec.~\ref{sec:NLeffects}.

As in Ref.~\cite{An:2015jdp}, we included systematic uncertainties concerning the flux, the detector efficiency (which are normalizations correlated among all bins,{  \it i.e. }$N_i\rightarrow \alpha N_i$) and a bin-to-bin uncorrelated shape uncertainty.
The shape uncertainty is simply introduced as an independent normalization for each bin in reconstructed energy, { \it i.e.} $N_i \rightarrow \alpha_i N_i$.

In the next section we will discuss in detail how some experimental issues can affect JUNO's ability to determine the neutrino mass ordering\footnote{For a verification of our simulation, see Appendix~\ref{app:compare}.}. 
We will concentrate  on the impact of  the real reactor core distribution, the inclusion of background events, the bin to bin flux uncertainty,  the number of  equal-size energy bins of data and the detector energy resolution.
We leave the discussion of the dependence on the true value of the neutrino oscillation parameters,  on the non-linearity of the detector energy response and on statistical fluctuations for later sections.

\section{ Mean (or Average) Determination of the neutrino mass ordering}
\label{sec:massordering}

In the following subsections we will discuss in which way the following quantities  affect the determination power of the neutrino mass ordering of the JUNO experiment:
\begin{enumerate}
 \item[A.] Effect of the reactor distribution and backgrounds,
 \item[B.] Effect of bin to bin flux uncertainties,
 \item[C.] Effect of varying the number of energy bins,
  \item[D.] Effect of varying the energy resolution.
\end{enumerate}

Unless otherwise stated, we generate fake data fixing the neutrino oscillation parameters as in Tab.~\ref{tab:oscparam} and assume the nominal values for the energy resolution, number of data bins and total exposure for JUNO given in Tab.~\ref{tab:sys}. 
\begin{table}[h]
\centering
  \catcode`?=\active \def?{\hphantom{0}}
    \begin{tabular}{|c|c||c|c|}
    \hline
    Quantity &~ Nominal Value ~ & ~Lowest Value ~& ~Largest Value
    \\
    \hline\hline
    $\epsilon$   (resolution @ 1MeV) & 3.0\% & 2.9\% & 3.1\%\\  
    b2b & 1\% & 0\% & 3\% \\
    $ \sigma_\text{bias}$ & 0.7\% & 0\% & no penalty\\ 
    number of bins & 200 & 100 & 300  \\
    exposure  (years) @ 26.6\,GW$_{\text{th}}$ ~& 8 & 2 & 16 \\
    \hline
    \end{tabular}
    \caption{Nominal values, as well as lowest and largest values, assumed in this paper for the JUNO energy resolution, systematic uncertainties (b2b=bin to bin and 
    the energy scale bias $\sigma_\text{bias}$), number of energy data bins  and exposure. One year is 300 days of live time.}
    \label{tab:sys} 
\end{table}


\subsection{Effect of the Reactor Distribution and Backgrounds}
\label{sec:react_dist}

\begin{figure}[h]
\centering
   \includegraphics[width=0.65\textwidth]{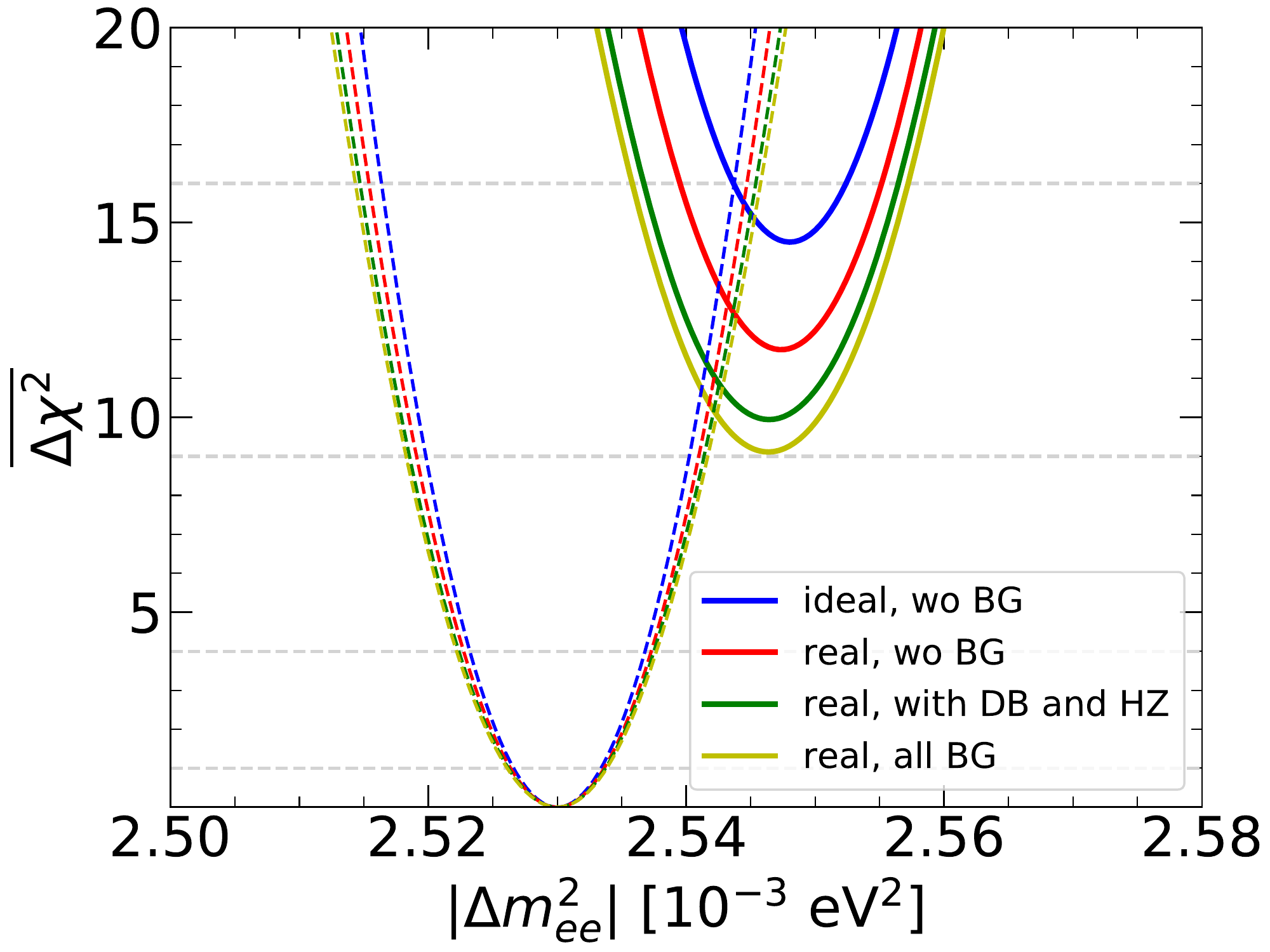}
    \caption{The  effects of the real reactor core-detector  baseline  distribution as well as  of the two types of backgrounds: 
     from the distant reactors Daya Bay (DB) and Huizhou (HZ) as well as  from  other sources (accidental, cosmogenic, etc.). 
     Going from the ideal distribution (all  cores  at  52.5 km) with no backgrounds (blue) to the real distribution (Table \ref{tab:distr}) with all backgrounds (dark yellow) the 
    $\overline{\chi^2}_\text{min}[\text{IO}]$ goes from 14.5 to  9.1, i.e. a reduction of more than 5 units.  Here ``wo" is abreviation for ``without".}
  \label{fig:distr}
\end{figure}

The real position of the reactor cores and background events are expected to impact JUNO's sensitivity.
Fig. \ref{fig:distr} shows the reduction in  $\overline{\Delta \chi^2}$ as one goes from the {\it ideal} reactor core-detector 
disposition (all cores at 52.5 km) with no backgrounds included to the real reactor core-detector baseline distribution given in Table \ref{tab:distr} with  all backgrounds taken into account.
The blue lines, labeled ``ideal, wo BG'', are the same as on the right panel of Fig.~\ref{fig:spectra}. 

There are two types of background events at JUNO: one from remote reactors (Daya Bay and Huizhou) and the other 
 includes accidental events, cosmogenic decays and geo-neutrinos. The first we compute, the latter we take from~\cite{An:2015jdp}.
  
Notice the  $ \overline{\chi^2}_\text{min}[\text{IO}]$ goes from  14.5 (ideal, wo BG) down to 9.1 (real, all BG), a decrease of more than 5 units.
The real core positioning alone causes a reduction in sensitivity of 2.8 and the background events an extra 2.6  (1.8 from DB and HZ).
We use the real baseline distribution and include all backgrounds in the rest of this paper.

\begin{table}[h]
\centering
\begin{tabular}{|c||c|c|c|c|c|c|c|c||c|c||}
\hline
Reactor & YJ-C1 &  YJ-C2 &  YJ-C3 &  YJ-C4 &  YJ-C5 &  YJ-C6 & TS-C1 & TS-C2 & ~DB~ & HZ \\
\hline
Power (GW$_{\text{th}}$) &   2.9 & 2.9 & 2.9 & 2.9 & 2.9 & 2.9 & 4.6  &  4.6 & 17.4  &  17.4 \\
Baseline (km) 
& 52.74 & 52.82 & 52.41 & 52.49 & 52.11 & 52.19& 52.77 & 52.64  & 215 & 265  \\
\hline
\end{tabular}
\caption{ The thermal power and core-detector baselines for the Yangjiang (YJ) and Taishan (TS) reactors, see \cite{Abusleme:2021zrw}. The total power is 26.6 GW$_{\text{th}}$.
The remote reactors Daya Bay (DB) and Huizhou (HZ) produce background events for the neutrino mass ordering.}
\label{tab:distr}
\end{table}


\subsection{Effect of bin to bin Flux Uncertainties}
\label{sec:flux_det}

There is uncertainty related to the exact shape of the reactor $\bar \nu_e$ flux, inherent to the flux calculation. This uncorrelated  bin to bin (b2b) shape uncertainty is included in our analysis by varying each predicted event bin with a certain penalty. The primary purpose of the TAO near detector is to reduce this bin to bin shape uncertainty, see \cite{Abusleme:2020bzt}.

\begin{figure}[bth]
 \centering
     \includegraphics[width=0.65\textwidth]{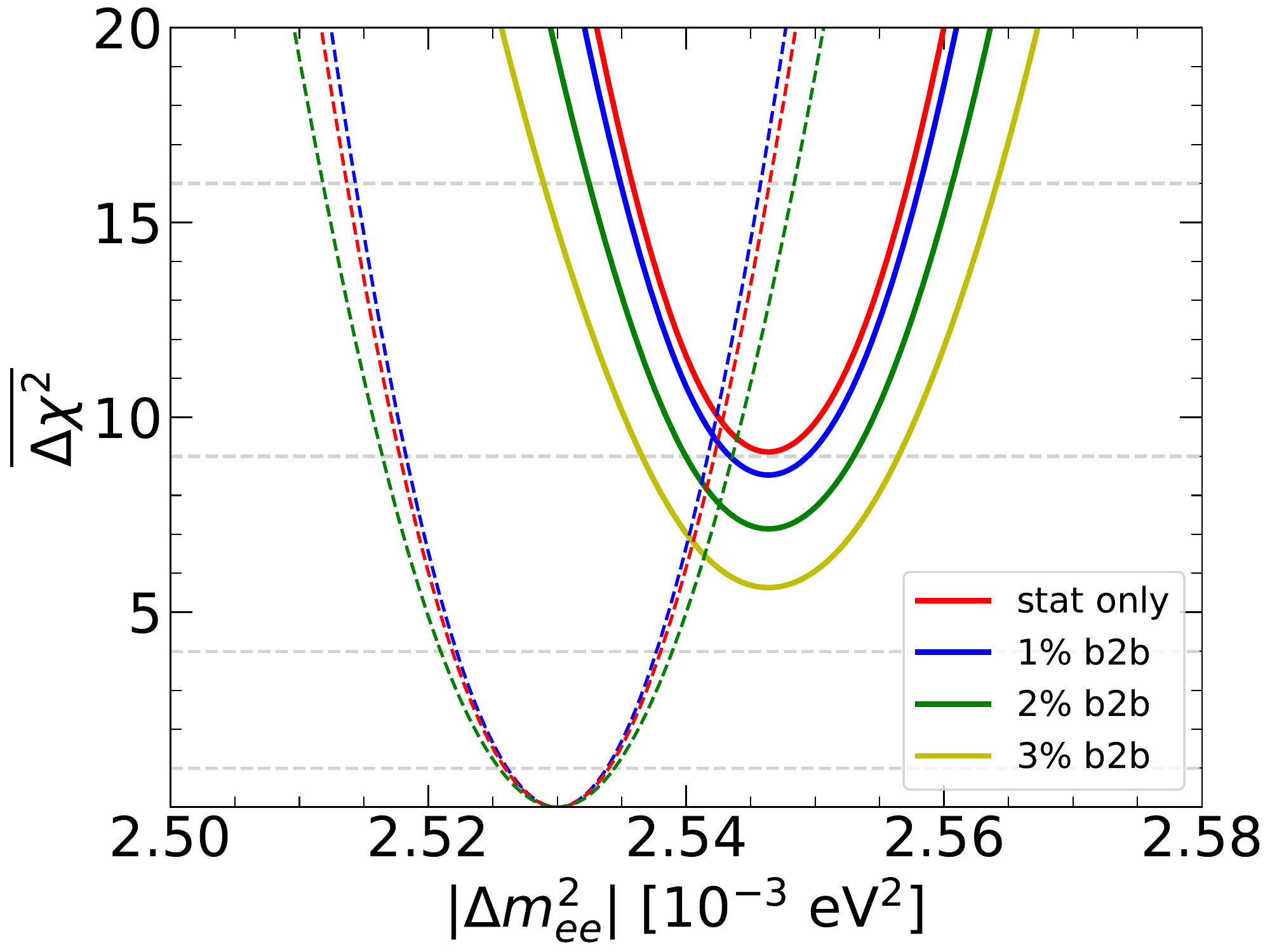}
    \caption{
    The effect of the bin to bin (b2b) systematic uncertainty on the  $ \overline{\chi^2}$.  The real distribution of reactors  is used  and all backgrounds are included.
     A 1\% b2b uncertainty is expected to be achieved with the TAO near detector \cite{Abusleme:2020bzt}. }
  \label{fig:b2b}
\end{figure}

The effect of this systematic  bias is shown in Fig.~\ref{fig:b2b}. The lines labeled ``stat only'' is the same as the one labeled  ``real, all BG'' in Fig. \ref{fig:distr}.  
We find $\overline{\chi^2}_{\text{min}}[\text{IO}]=8.5, 7.1$ and  $5.6$, respectively, for 1\%, 2\% and 3\%. 
When the b2b systematic uncertainty is not included, we recall,  $\overline{\chi^2}_{\text{min}}[\text{IO}] = 9.1$.
So if the shape uncertainty is  close to $1\%$ (the nominal value), the sensitivity to the neutrino mass ordering is barely affected.
However, for 2\% and 3\% we see a clear loss in sensitivity. 
This is because increasing  the uncorrelated uncertainty for each bin, makes it easier to shift from a NO spectrum into an IO one 
and vice versa.
We use 1\% b2b in the rest of the paper.

\newpage

\subsection{Effect of varying the number of Energy Bins}
\label{sec:bins}

\begin{figure}[h]
  \centering
   \includegraphics[width=0.65\textwidth]{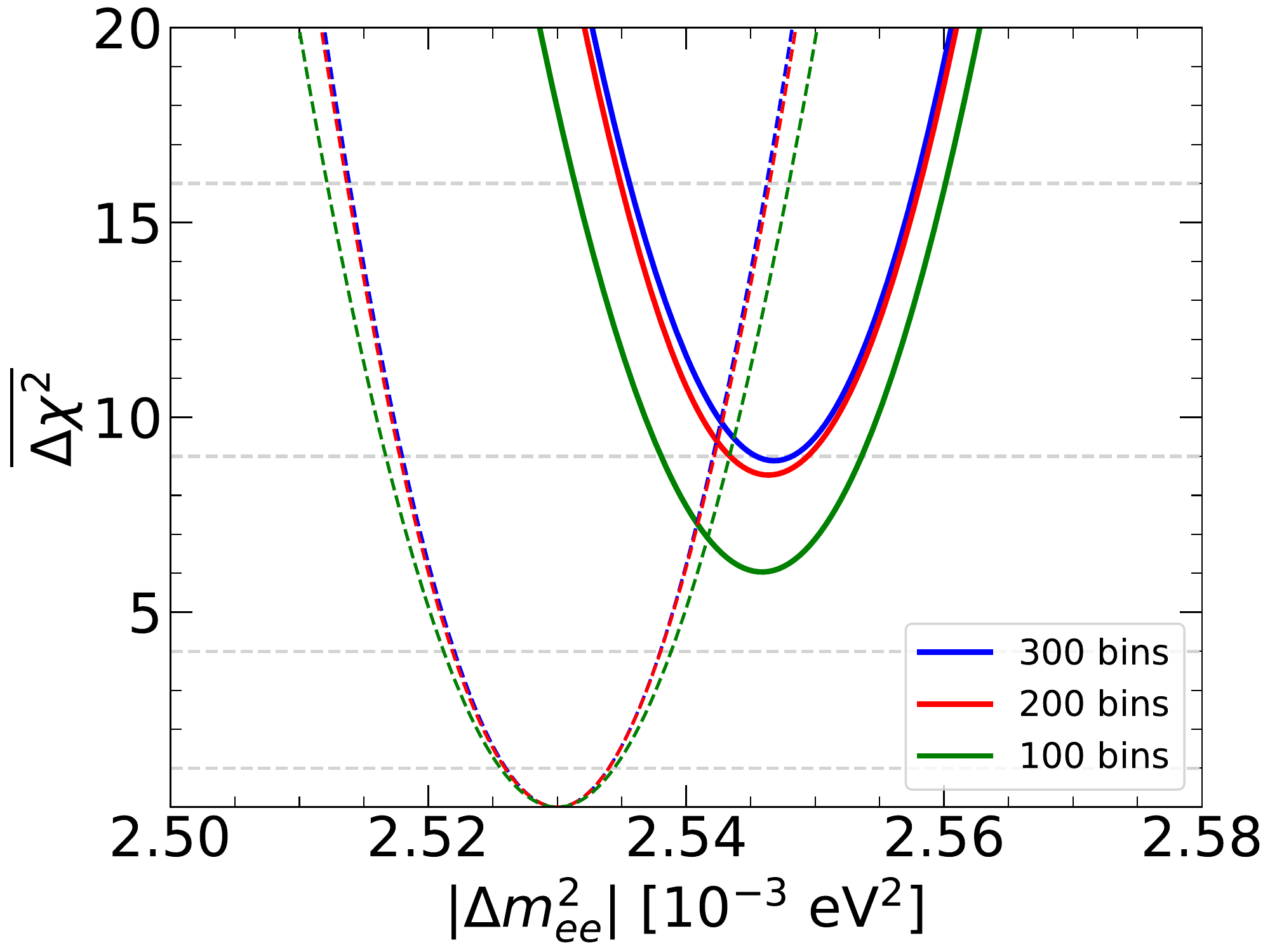}
       \caption{
       The effect of varying the number of { neutrino energy} binsin the range [1.8, 8.0]~MeV on the  $\overline{ \chi^2}$. 
        Below 200 bins the  $\overline{\chi^2}_\text{min}[\text{IO}]$ decreases substantially whereas above 200 the   increase is 
        marginal. 
       }
  \label{fig:bin}
\end{figure}

Here we examine the impact on $\overline{\chi^2}$ of changing the size of the neutrino energy bins in the range [1.8, 8.0] MeV. In Fig.~\ref{fig:bin}, we show the result obtained from varying the number of energy bins.
We obtain $\overline{\chi^2}_\text{min}[\text{IO}]= 6.0, 8.5$ and 8.9, respectively, for 100, 200 and 300 bins.
So increasing the number of bins above 200  causes a marginal improvement, whereas lowering the number of bins below 
200 reduces the significance of the neutrino mass ordering determination. 
The background per bin from Ref.~\cite{An:2015jdp} is re-scaled as we vary the number of bins. The red lines in Fig.~\ref{fig:bin} (200 bins) are the same as the blue lines in Fig.~\ref{fig:b2b} (1\% b2b). We always use 200 bins elsewhere in this paper.

\newpage

\subsection{Effect of varying the  Energy Resolution}
\label{sec:res}

\begin{figure}[h]
  \centering
  \includegraphics[width=0.65\textwidth]{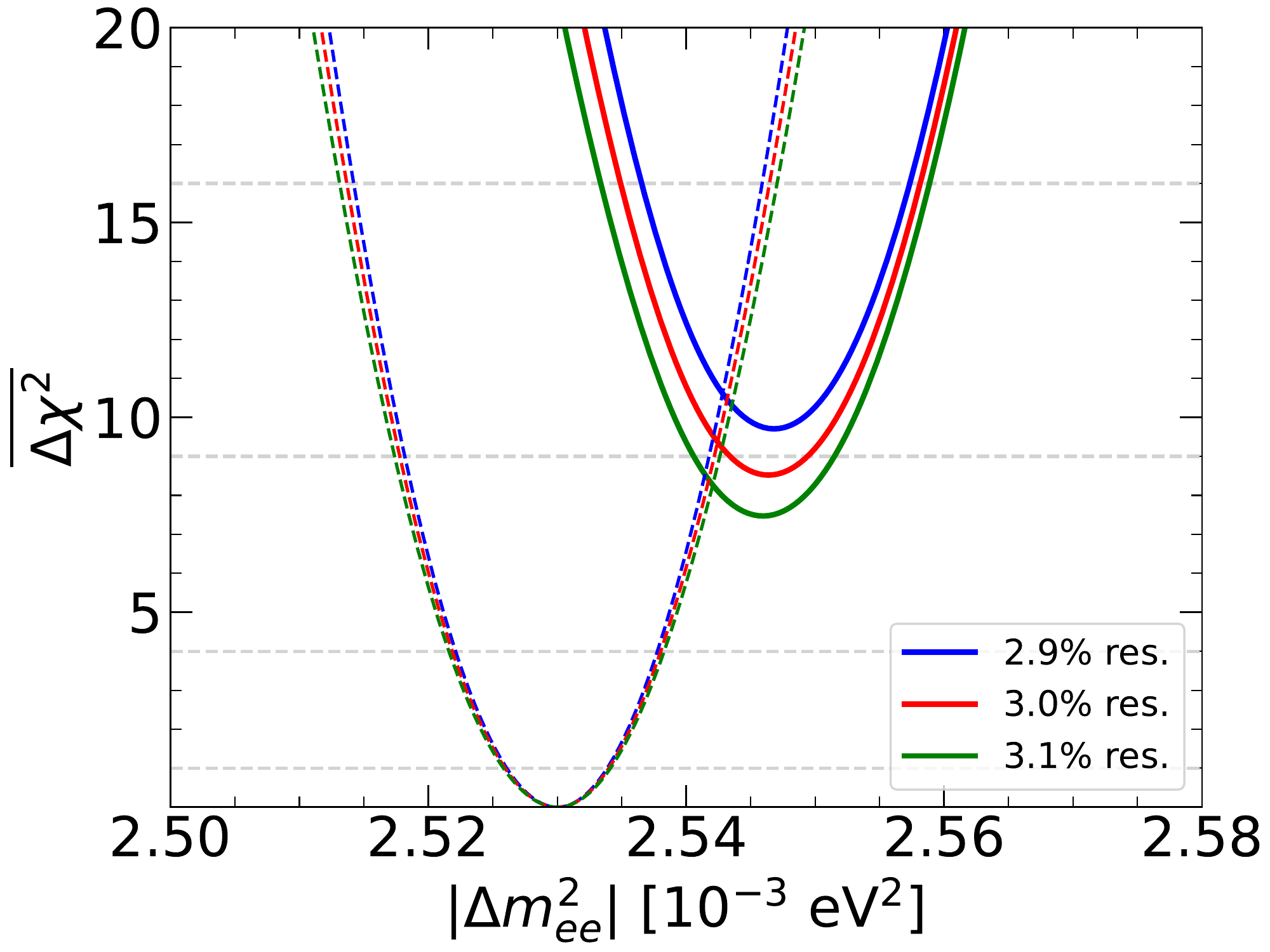}
       \caption{Here we show the effect of varying the detector  energy away from the nominal  3.0\%. A 0.1\% reduction (increase) in this resolution increases (decreases)  the   $\overline{\chi^2}_\text{min}[\text{IO}]$ by approximately 1 unit.}
  \label{fig:res}
\end{figure}

Next we consider variations of the  detector energy resolution. 
In this section we assume, that this number can be slightly better and slightly worse than the nominal 3.0\%.
Small variations of the resolution have large impacts on the determination of the neutrino mass ordering, as shown in Fig.~\ref{fig:res}.
The red line corresponds to the nominal energy resolution of 3.0$\%$. The blue and green lines are obtained for 2.9$\%$ and $3.1\%$, respectively, with corresponding  $\chi^2_\text{min}[\text{IO}]=9.7$  and 7.5. Clearly $\chi^2_\text{min}[\text{IO}]$ is quite sensitive to the exact value of the resolution that will be achieved by JUNO. 
Therefore even a small improvement on the energy resolution would have a sizable impact on the determination potential of the neutrino mass ordering. 
However, it appears challenging for JUNO to  reach an energy resolution even slightly better than 3.0\%, see \cite{Abusleme:2020lur}.
Note the red lines in  Fig.~\ref{fig:res}  (3.0\% res.) are also the same as the  blue lines in Fig.~\ref{fig:b2b} (1\% b2b). We always use 3.0\% resolution elsewhere  in this paper.

\newpage

\section{ Effect of varying the true values of the Neutrino Oscillation Parameters }
\label{sec:osc_effect}

In this section we explore how varying the true values of the neutrino oscillation parameters improves or reduces the prospects for JUNO's determination of the neutrino mass ordering.  We first consider the variation of  single parameters with the others held fixed and then consider the correlations varying both  $\Delta m^2_{21}$ and $\sin^2 \theta_{12}$ with $\Delta m^2_{ee}$ and $\sin^2 \theta_{13}$ held fixed and vice versa.

\begin{figure}[h]
  \centering
  \includegraphics[width=0.8\textwidth]{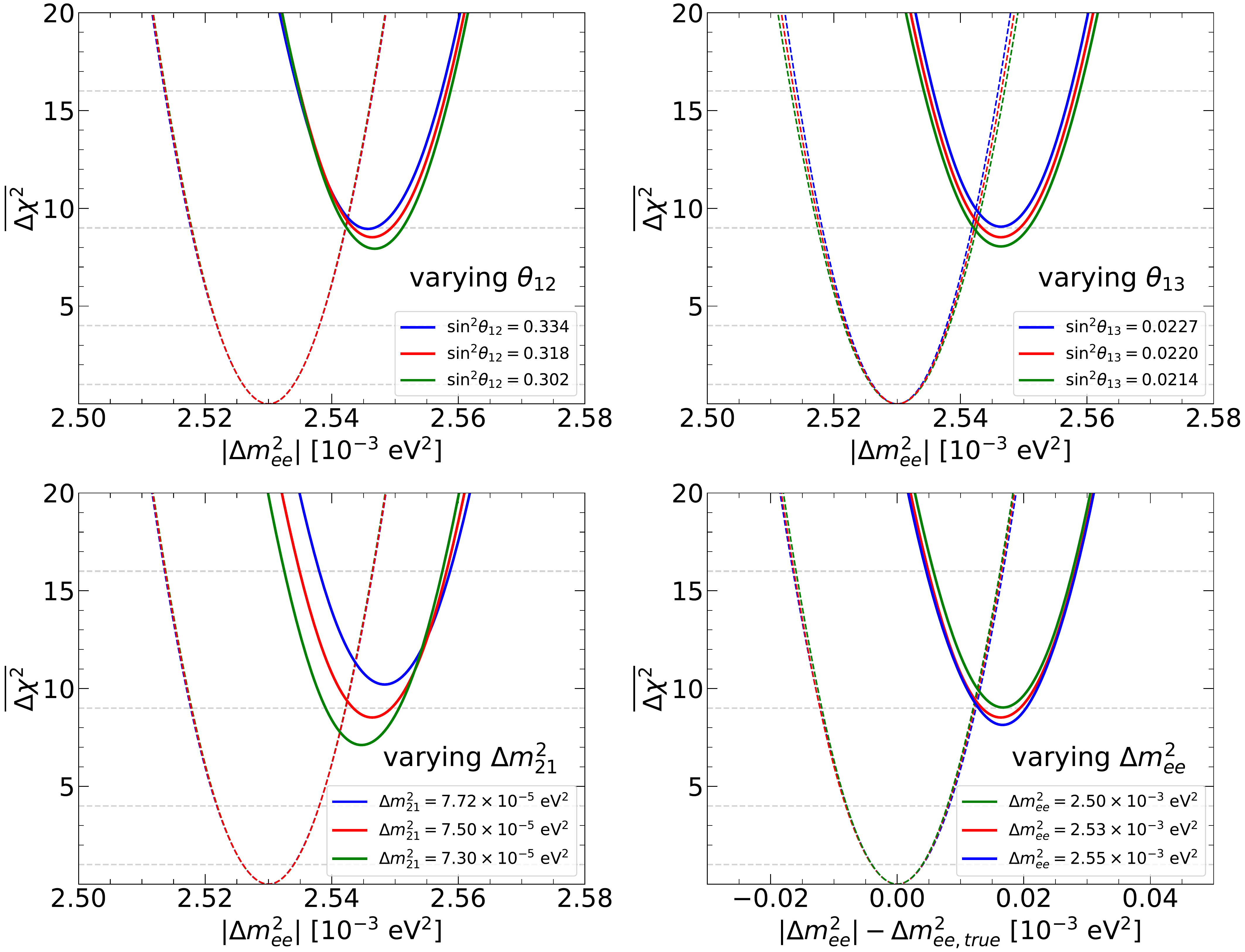}
    \caption{ How the
  $ \overline{\chi^2}$  dependency on the true values of the neutrino oscillation parameters can impact    
   the neutrino mass ordering determination. 
    The  curves for the global best fit value (red)  and the curves for a value 1$\sigma$ above (below) from the  global best fit  are shown 
    in blue (green), according to Tab.~\ref{tab:oscparam}.
     Only the labeled parameter is varied in each plot, the others are held at their best fit values. 
Here we use the nominal values  for resolution, b2b systematics, number of energy bins and exposure  given in Tab.~\ref{tab:sys} and include all backgrounds.}
  \label{fig:osc_par_vary}
\end{figure}

We start by creating fake data sets using the upper and lower 1$\sigma$ bounds obtained in Ref.~\cite{deSalas:2020pgw}  (see Tab.~\ref{tab:oscparam}), always for one parameter at the time. 
The result of these analyses is shown in Fig.~\ref{fig:osc_par_vary}, where in each panel we vary one of the parameters as indicated.
 Here again solid (dashed) lines are  used for IO (NO).
As can be seen, changes in any of the oscillation parameters can have large effects on the determination power of the neutrino mass ordering. 
Especially remarkable is the effect of a smaller $\Delta m_{21}^2$, which shifts  $ \overline {\chi^2}_\text{min}[\text{IO}]$ from 8.5 to 7.1. 
Note that the best fit value obtained from the analysis of solar neutrino data from Super-K~\cite{yasuhiro_nakajima_2020_4134680} is even smaller than the one considered here and therefore the determination would then be even more difficult.
On the other hand side, a larger value of the solar mass splitting improves significantly the determination of the mass ordering. 
In this case we obtain $ \overline {\chi^2}_\text{min}[\text{IO}]=10.2$.
The effect of the other parameters is not as pronounced as in the case of the solar mass splitting, but still  appreciable: $\Delta m^2_{ee}$/$\sin^2\theta_{13}$/ $\sin^2\theta_{12}$, within 1$\sigma$ of their current best fit value, can move $ \overline {\chi^2}_\text{min}[\text{IO}]$ by approximately $\pm $ 0.5.

\newpage

\begin{figure}[htb]
  \centering
   \includegraphics[width=0.49\textwidth]{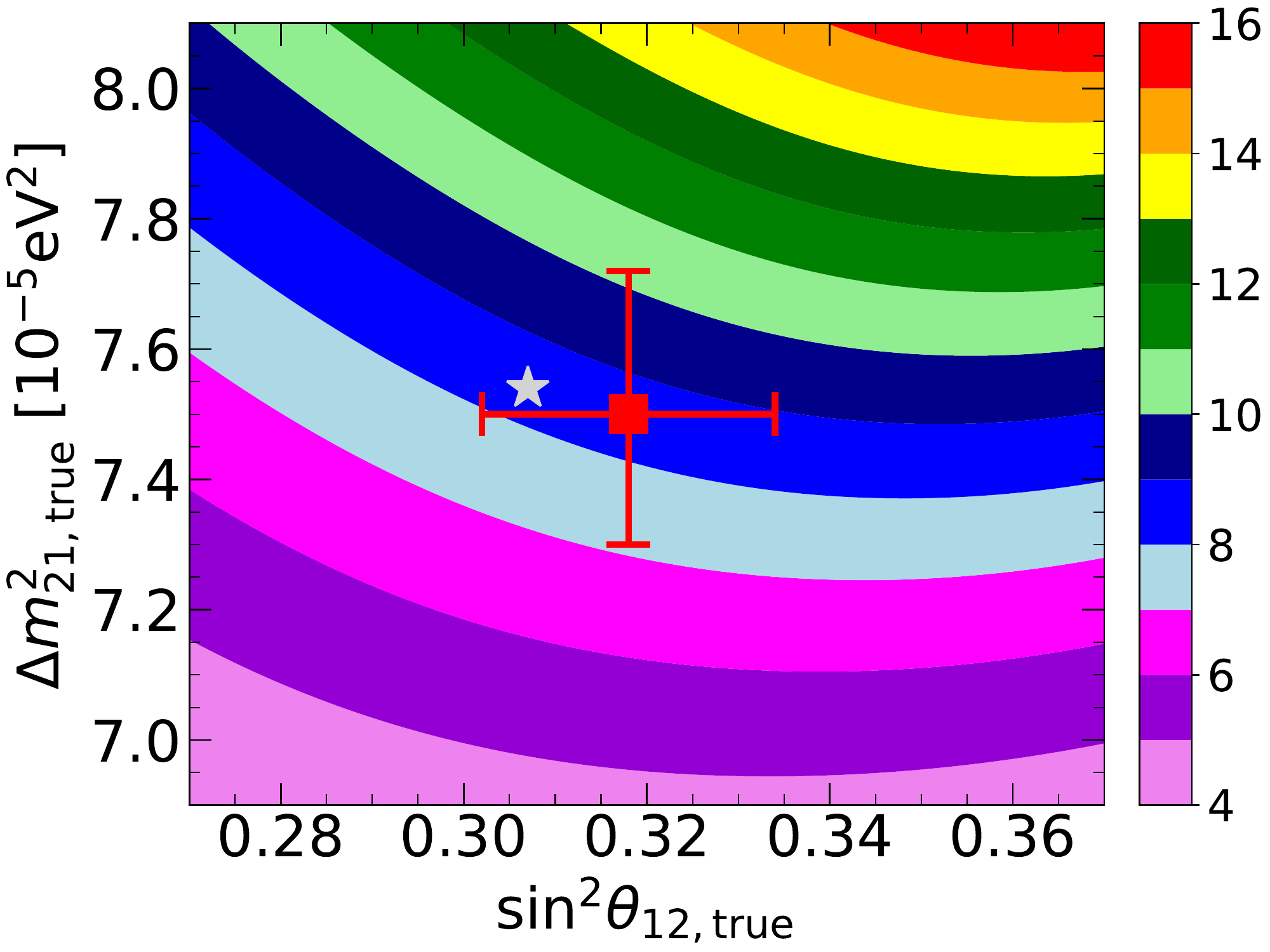}
     \includegraphics[width=0.49\textwidth]{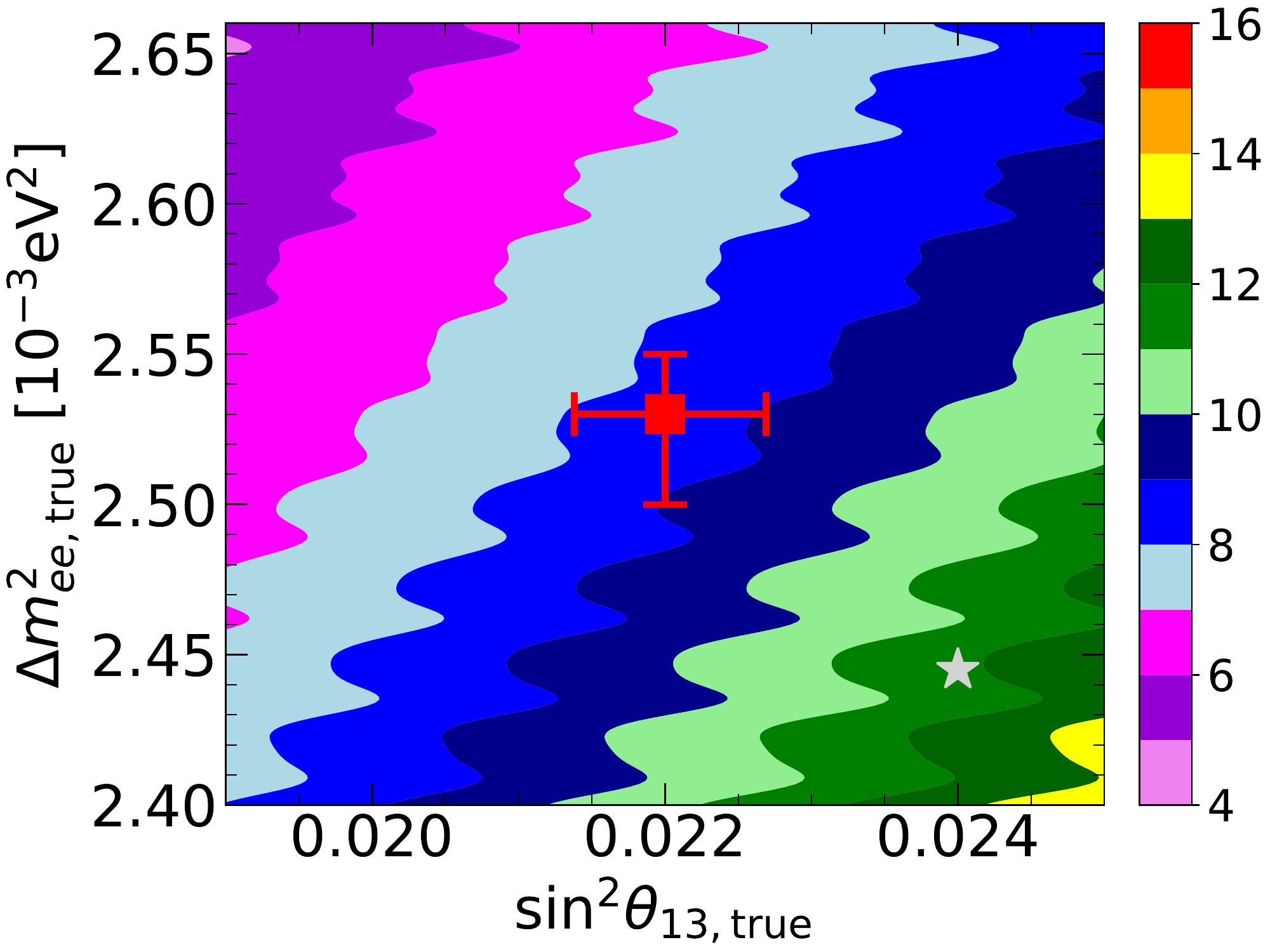}
    \caption{Contours of  $\overline{ \chi^2}_{\text{min}}[\text{IO}]$ as the oscillation parameters are varied: left panel varying ($\sin^2 \theta_{12}$, $\Delta m^2_{21}$) holding ($\sin^2 \theta_{13}$, $\Delta m^2_{ee}$) fixed at  their best fit values, right panel varying ($\sin^2 \theta_{13}$, $\Delta m^2_{ee}$) holding ($\sin^2 \theta_{12}$, $\Delta m^2_{21}$) fixed  at  their best fit values. 
 The red cross is the current best fit point whereas the gray star is the value of the parameters used in \cite{Li:2013zyd}  (also in \cite{An:2015jdp}).  Even for  
 a variation of about 3$\sigma$ around the best fit values of Tab.~\ref{tab:oscparam}, we see substantial change in the $\overline{\chi^2}_{\text{min}}[\text{IO}]$. 
 Here we use the nominal values  for resolution, b2b systematics, number of energy bins and exposure  given in Tab.~\ref{tab:sys} and include all backgrounds. }
  \label{fig:osc-par_iso}
\end{figure}

In Fig. \ref{fig:osc-par_iso} we show the correlated variation of the  $ \overline{\chi^2}_{\text{min}}[\text{IO}]$ as  a function of  ($\sin^2 \theta_{12}$, $\Delta m^2_{21}$) holding ($\sin^2 \theta_{13}$, $\Delta m^2_{ee}$) fixed as well as a function of  ($\sin^2 \theta_{13}$, $\Delta m^2_{ee}$) holding ($\sin^2 \theta_{12}$, $\Delta m^2_{21}$) fixed. Even varying these parameters within 3$\sigma$ of their current best fit, there are very significant changes to the $ \overline{\chi^2}_{\text{min}}[\text{IO}]$ contour plots.  This implies that JUNO's prospect for the determination of the neutrino mass ordering could be  improved or  weakened by Nature's choice for the true values of these oscillation parameters.  The values that were used in \cite{Li:2013zyd}  (also in \cite{An:2015jdp}) are shown by the gray stars in these figures.

\newpage

\section{Non-linear detector energy response}
\label{sec:NLeffects}

In a liquid scintillator detector, the true prompt  energy,  $E_p$, (positron energy plus $m_e$)  is not a linear function of  to the visible energy, $ E^\text{vis}$, in the detector. 
The main energy-dependent effects are the intrinsic non-linearity related to the light emitting mechanisms (scintillation and Cherenkov emission) and instrumental non-linearities.
The non-linear detector response can be modeled by a four parameter function~\cite{An:2013zwz,An:2016ses,Abusleme:2020lur} which relates the true prompt energy to the visible detector energy according to 
\begin{align}
 E_p &= \frac{E^\text{vis} }{f_\text{NL} (a_1,a_2,a_3,a_4; E_p)} \quad 
{\rm where} \quad  f_\text{NL} (a_1,a_2,a_3,a_4;E_p)  \equiv \frac{a_1 + a_2 \,(E_p/\text{MeV})}{1 + a_3 \, \exp\left(-a_4 \, (E_p/\text{MeV})\right)}\,,
 \label{eq:ratio0}
\end{align}
and the coefficients $(a_1,a_2,a_3,a_4)$ are determined by prompt energy calibration techniques.  We use the prompt energy scale calibration curve shown in Fig.~1 of  Ref.~\cite{Abusleme:2020lur}, which can be well described by the $f_\text{NL}$ given in Eq.~\eqref{eq:ratio0} with 

$$\bar{a}_1=1.049, \quad \bar{a}_2=2.062\times10^{-4}, \quad \bar{a}_3=9.624\times10^{-2}, \quad \bar{a}_4=1.184\,.$$
Then the true neutrino energy, $E$, is then constructed by $E=E_p+\Delta M$.

To allow for deviations from this calibration, we use in our simulation the reconstructed prompt energy, $E^{\,\prime}_p$, given by
\begin{align}
&\frac{E^{\,\prime}_p}{ E_p} =  \frac{f_\text{NL} (\bar{a}_1,\bar{a}_2,\bar{a}_3,\bar{a}_4; E_p) }{f_\text{NL} (a_1,a_2,a_3,a_4; E_p)}.
 \label{eq:ratio}
\end{align}
Note, with this definition $E^\text{vis}$ is held fixed as we change the $a_i$'s from their calibration values, $\bar{a}_i$.  In the simulation, we generate a distribution of $E_p$'s for the true mass ordering and use Eq.~\eqref{eq:ratio} to generate a distribution of  $E^{\,\prime}_p$'s for the test mass ordering\footnote{For the neutrino energy, the  equivalent expression is
 $$
\frac{E^{\, \prime}}{E} =  \frac{f_\text{NL} (\bar{a}_1,\bar{a}_2,\bar{a}_3,\bar{a}_4; E-\Delta M) }{f_\text{NL} (a_1,a_2,a_3,a_4; E-\Delta M)}
 \left(1-\frac{\Delta M}{E}\right )+\frac{\Delta M}{E} 
\,.
$$.
}.

\begin{figure}[tb]
  \centering
   \includegraphics[width=0.49\textwidth]{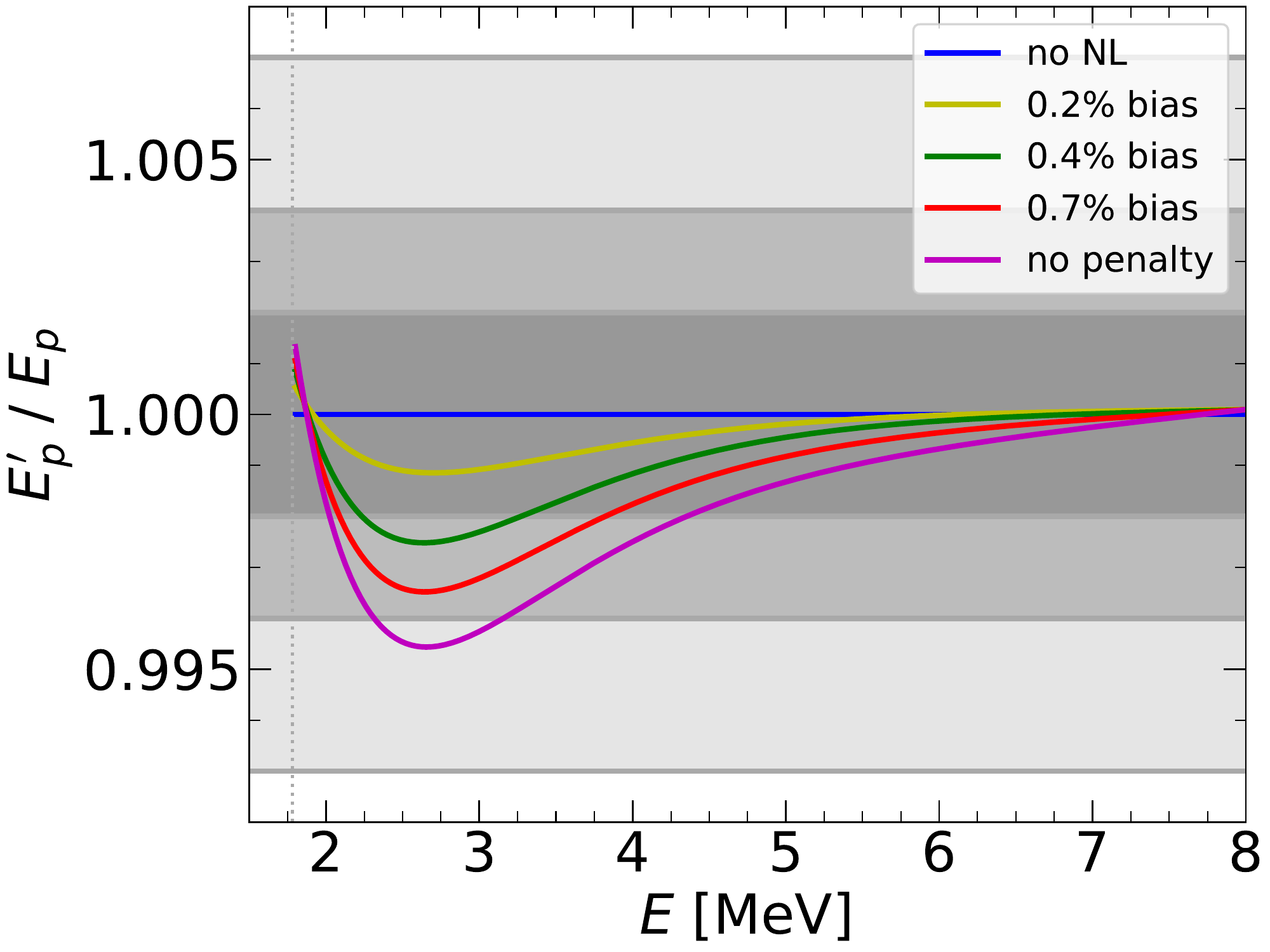}
     \includegraphics[width=0.49\textwidth]{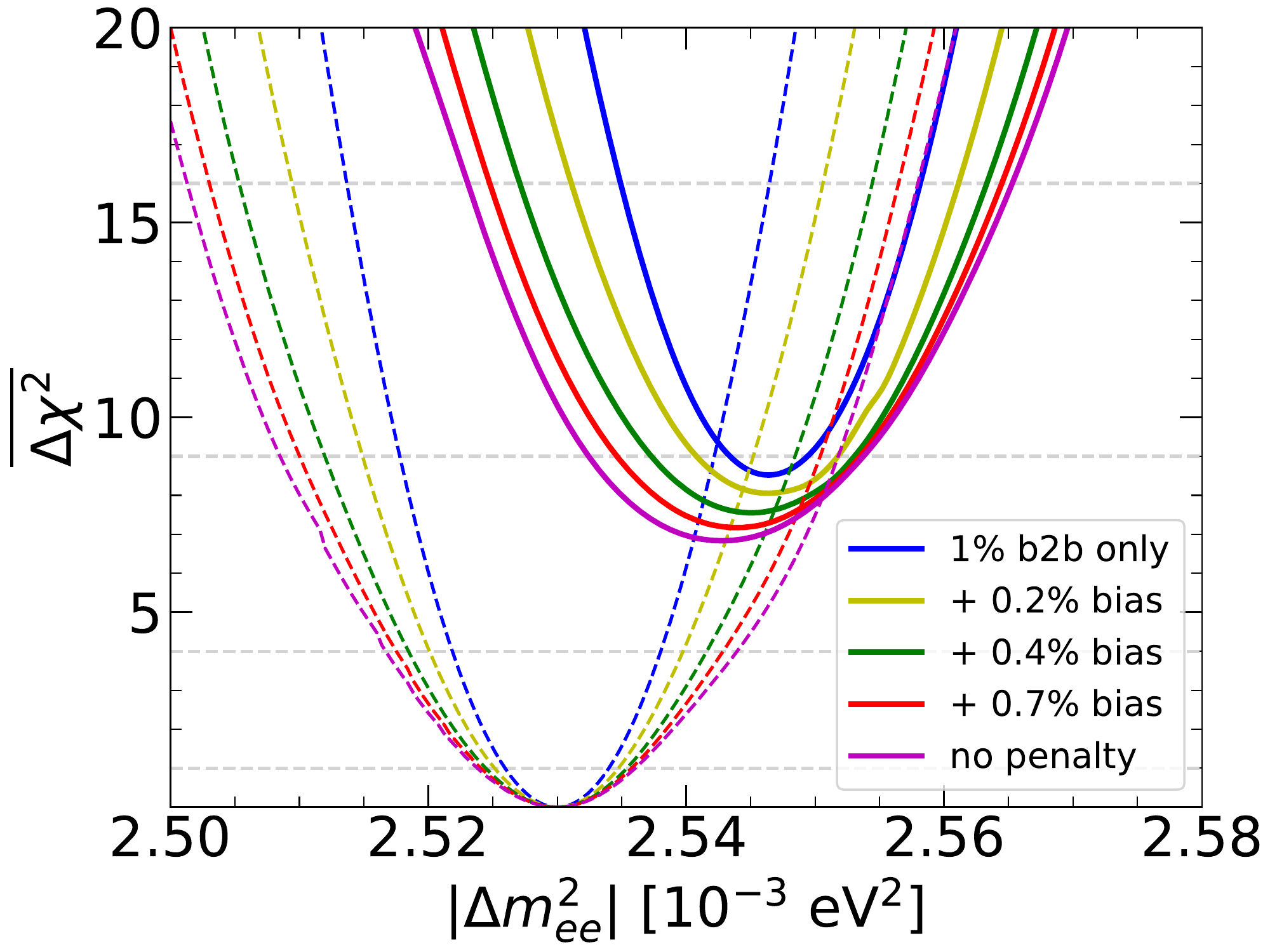}
    \caption{On the  left panel we show the ratio between the reconstructed prompt energy $E^\prime_p$ and the true prompt energy $E_p$ as a function of  the 
    neutrino  energy $E,$ 
    for $\sigma_\text{bias}=0.2\%$ (yellow), $0.4\%$ (green) and $0.7\%$ (red) and no penalty (magenta) for the  best fit  to IO for the NO spectra. In blue we show the line for perfect reconstruction (no NL) as a reference.
On the right panel   we show  the changes  to $\overline{\chi^2}$ caused by the addition of the corresponding $\chi^2_\text{NL}$.}
  \label{fig:nlbias}
\end{figure}

The allowed range of the $a_i$'s is constrained by   including a penalty term for the derivation of
$$\frac {|E^{\,\prime}_p-E_p|}{E_p} \, ,$$
when fitting the  simulated spectra to the test mass ordering. 
Explicitly, we allow the $a_i$'s to vary from their calibration values and then penalize  the fit by using the simplified $ \chi^2_\text{NL}$ defined, as in Ref.~\cite{Capozzi:2015bpa}, as
\begin{equation} 
\chi^2_\text{NL}= \max_{E_p}  \left( \frac{f_\text{NL} (\bar{a}_1,\bar{a}_2,\bar{a}_3,\bar{a}_4;E_p)}{f_\text{NL} (a_1,a_2,a_3,a_4;E_p)} -1 \right)^2\biggr/ (\sigma_\text{bias})^2\, ,
 \label{eq:bias}
\end{equation}
where $\{a_i, i=1,...,4\}$ are the best fit  of these parameters for the test mass ordering spectra,  $\sigma_\text{bias}$ is the uncertainty on the energy scale,
$ \displaystyle \max_{E_p}$ indicates that we take only the maximal difference which  happens at $E \sim 2.75$~MeV (see Fig.~\ref{fig:nlbias}).
We consider the following sizes for  the bias, $\sigma_\text{bias}=0.0, 0.2, 0.4, 0.7\%$, as well as no penalty, {\it i.e.} ~~$\chi^2_\text{NL}=0$.
Using Fig.~2 of  Ref.~\cite{Abusleme:2020lur}, we see that JUNO expects an approximately energy independent systematic uncertainty  on the energy scale of about 0.7\%, mainly due to instrumental non-linearity and position dependent effects. Therefore, $\sigma_\text{bias}=0.7\%$ is our nominal value from here on.

On the left panel of Fig.~\ref{fig:nlbias} we show the ratio $E_p'/E_p$ as a function of $E$ for the corresponding $f_\text{NL}$ coefficients listed in Tab.~\ref{tab:bais} 
obtained for the best fit to IO of the NO input spectra.
On the right panel we see the effect of the uncertainty  on the energy scale on $\overline{ \chi^2}$. In particular, as the uncertainty increases $\overline{ \chi^2}_\text{min}[\rm IO]$  goes from 8.5 (no NL effect) down to 8.0 (0.2\%), 7.5 (0.4\%) and 7.2 (0.7\%). Note that if we introduce the non-linearity shift with no  
penalty $\overline{ \chi^2}_\text{min}[\rm IO]=6.8$. Even with the nominal 0.7\% bias, this is a significant effect, reducing $\overline{ \chi^2}_\text{min}[\rm IO]$ by more than 1 unit   (8.5 to 7.2) and in this manner further lowering the mass ordering discrimination power.

We also observe that the precision on the determination of $\vert\Delta m^2_{ee}\vert$ is  notably degraded when the non-linearity in the energy scale is included.
In addition, the best fit value for $\vert\Delta m^2_{ee}\vert[\rm IO]$ moves slightly toward   the best fit value for 
$\vert\Delta m^2_{ee}\vert[\rm NO]$. This means that the fit for IO adjusts the $f_\text{NL}$ coefficients in order to get a value for  $\vert\Delta m^2_{ee}\vert[\rm IO]$ closer to the input value of $\vert\Delta m^2_{ee}\vert[\rm NO]$.

\begin{table}[h]
\centering
\begin{tabular}{|c||c|c|c|c||}
\hline
           & $a_1$ &  ~~$ a_2  \times 10^{4} $~~  &  ~~$ a_3\times 10^{1}$   ~~&  $a_4$\\
\hline
Calibration  & \, 1.049  \,  &  \,  2.062 \, & \,  0.9624 \,    &  \,   1.184 \,  \\
0.2\% &  \,  1.049 \,  & \,  1.918 \, & \,  1.156 \, & \, 1.347 \,  \\
0.4\%    &   \, 1.049  \, &  \, 1.633 \, & \,  1.424 \, & \, 1.534  \, \\
0.7\%   &   \, 1.050 \,  & \,  0.474 \,   & \, 1.614 \,  & \, 1.627 \, \\
No Penalty    & \, 1.051  \,   &\, -1.148   \, & \, 1.840 \, & \, 1.716 \, \\
\hline
\end{tabular}
\caption{Values of the coefficients of the function $f_\text{NL}$ for the calibration and 0.2, 0.4, 0.7\% bias as well as no penaltly.  }
\label{tab:bais}
\end{table}

\newpage

\section{Fluctuations about the Mean for the neutrino mass ordering determination}
\label{sec:fluct}

It has been already pointed out that statistical fluctuations are important for JUNO, see for instance Ref.~\cite{Ge:2012wj} where they estimate the statistical uncertainty  on $\overline{\Delta \chi^2}$  by an analytical  expression and a Monte Carlo simulation. The calculation was performed just after the first measurement of $\sin^2 \theta_{13}$ by RENO and Daya Bay, under different detector resolution and systematic assumptions. It is timely to reevaluate this here.

\begin{figure}[t]
  \centering
  \includegraphics[width=0.55\textwidth]{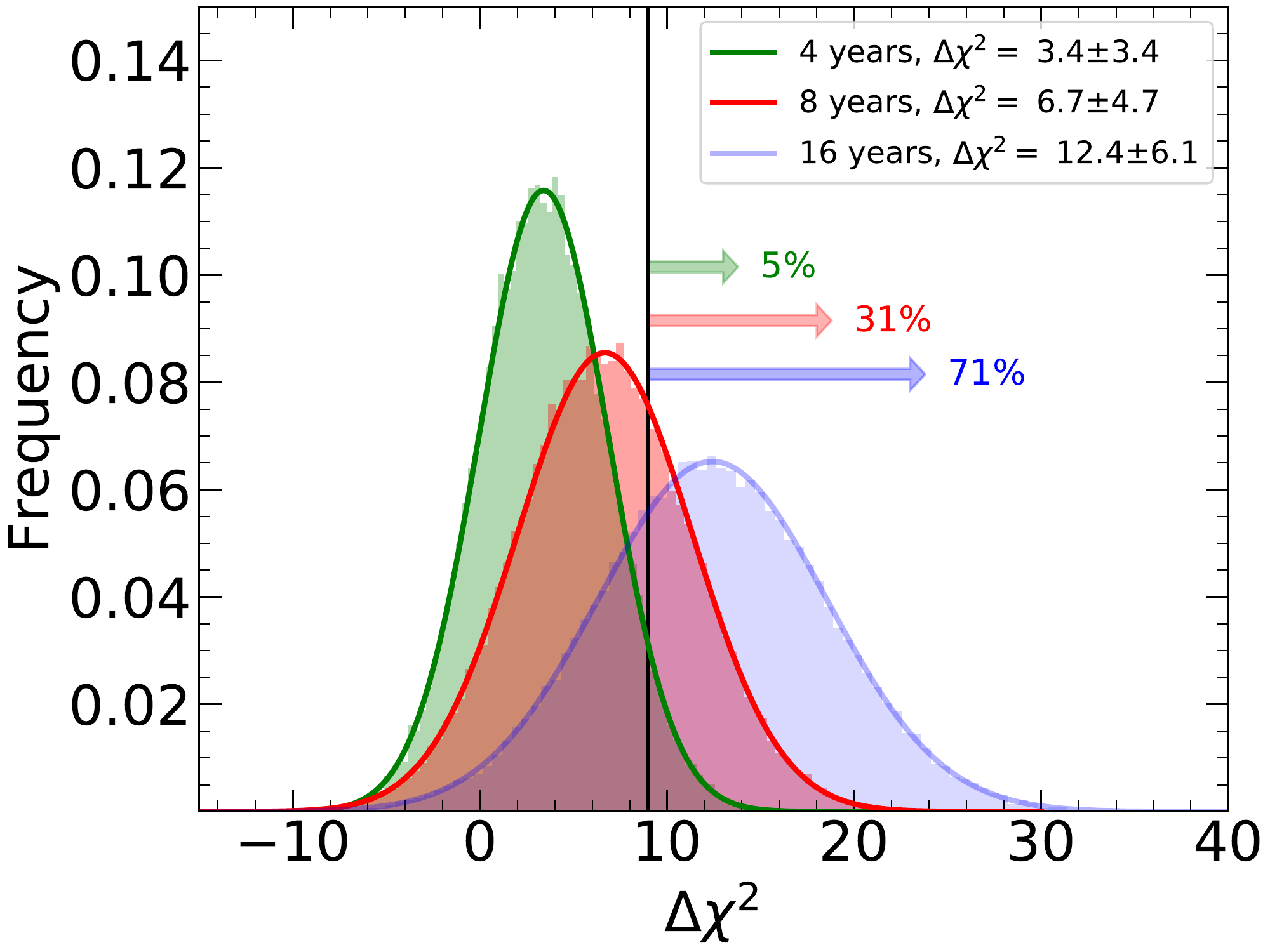}
    \caption{Distributions of the $\Delta\chi^2 \equiv \chi^2_\text{min}[\rm IO]- \chi^2_\text{min}[\rm NO]$ values obtained in the analyses of  60\,k trial pseudo-experiments where statistical fluctuations of the trial data have been taken into account for 
    three different exposures: 4 (green), 8 (red) and 16 (blue) years. We use the neutrino oscillation parameters at the values given in Tab.~\ref{tab:oscparam} and take into account the experimental nominal systematic uncertainties and energy resolution given in Tab.~\ref{tab:sys}. }
  \label{fig:MC}
\end{figure}

We have already shown in Fig.~\ref{fig:spectra} that the difference between the spectra for NO and IO is smaller than the statistical uncertainty in each bin. We consider here the effects of fluctuating the number of events in each bin. We evaluate the impact of this fluctuations on the mass ordering determination by performing a simulation of 60000 JUNO pseudo-experiments for each exposure and obtain the distributions given in Fig.~\ref{fig:MC}.
To generate this figure, we create a fake data set  \{$N^0_i, i=1,...,N_{\text{bins}}$\} using the neutrino oscillation parameters in Tab.~\ref{tab:oscparam}.
The fluctuated spectrum \{$N^f_i,i=1,...,N_{\text{bins}}$\} is generated by creating normal distributed random values around $N^0_i\pm\sqrt{N^0_i}$.
We analyze this fluctuated spectrum for NO and IO and add the corresponding $\Delta \chi^2 \equiv  \chi^2_\text{min}[\rm IO]- \chi^2_\text{min}[\rm NO]$ value to a histogram. 
Note that here, because of the statistical fluctuations, $\chi^2_\text{min}$[NO] is not necessarily zero, so $\Delta \chi^2<0$ means  $\chi^2_\text{min}$[NO]$>\chi^2_\text{min}$[IO], so the wrong mass ordering is selected in this case.

We use the  nominal values for the systematic uncertainties and energy resolution given in Tab.~\ref{tab:sys} for three exposures:
4, 8 and 16 years. The corresponding   $\Delta \chi^2$ distributions are shown in Fig.~\ref{fig:MC}. These distributions are Gaussian (as was proven analytically in Ref.~\cite{Blennow:2013oma}) with corresponding central values $\Delta\chi^2=3.4, 6.7$ and 12.4 and standard deviations 3.4, 4.7 and 6.1, respectively.
Our pseudo-experiments reveal that after 8 years in only 31\% of the trials JUNO can determine the neutrino mass ordering at the level of 3$\sigma$ or better. We also find that there is even a non negligible probability ($\sim$8\%) to obtain the  wrong mass ordering, {\it i.e.,}  $\Delta\chi^2<0$.
For a shorter (longer) exposure of 4 (16) years, $5\%$ $(71\%)$ of the pseudo-experiments rule out IO at 3$\sigma$ or more. In these cases in about 16\% (2\%) of the trials the IO is preferred.

\section{Combining JUNO with the Global Fit}
\label{sec:combined}

In the previous section we have shown that the significant impact of statistical fluctuations on top of the detector systematic effects, can make it very challenging for JUNO by itself to determine at 3$\sigma$ or more the neutrino mass ordering even after 16 years.  
However, as was shown in \cite{Nunokawa:2005nx}, muon disappearance experiments measure\footnote{In fact, there is a small correction to this definition whose leading term depends on $\cos \delta \sin \theta_{13} \sin 2\theta_{12}  \tan \theta_{23} \Delta m^2_{21}$ whose magnitude is less than $ 10^{-5}$ eV$^2$. This term is included in all numerical calculations.} 
\begin{equation}
\Delta m^2_{\mu\mu}  \equiv \sin^2 \theta_{12}  \Delta m^2_{31}+\cos^2\theta_{12}  \Delta m^2_{32} \, ,
\end{equation}
whose relationship to  $\vert \Delta m^2_{ee}\vert$ is given by
\begin{equation}
 \vert \Delta m^2_{ee}\vert = \vert \Delta m^2_{\mu\mu}  \vert \pm \cos 2 \theta_{12} \Delta m^2_{21} \,,
\end{equation}
the positive (minus) sign is for NO (IO). Therefore, by using muon disappearance measurements we have a constraint on the allowed $\vert \Delta m^2_{ee} \vert$'s for the two mass orderings,
\begin{equation}
\vert \Delta m^2_{ee} \vert\,[{\rm NO}]- \vert \Delta m^2_{ee} \vert\,[{\rm IO}] = 2 \cos 2 \theta_{12} \Delta m^2_{21} \approx 0.06 \times 10^{-3}\, \text{eV}^2 \, ,
\end{equation}
i.e. $\vert\Delta m^2_{ ee}\vert\,[{\rm IO}]$ is  2.4\% {\it smaller} than $\vert\Delta m^2_{ ee}\vert\,[{\rm NO}]$. 
Whereas, because of the phase advance (NO) or retardation (IO) given in Eq.~\eqref{eq:phiodot}, the medium baseline reactor experiments give  $\vert\Delta m^2_{ee}\vert\,[{\rm IO}]$ about 0.7\% {\it larger} than $\vert\Delta m^2_{ee}\vert\,[{\rm NO}]$. Of course, the measurement uncertainty on  $ \vert \Delta m^2_{\mu\mu}  \vert $ must be smaller than this 3.1\% difference for this measurement to impact the confidence level at which the false mass ordering is eliminated. 
The short baseline reactor experiments, Daya Bay and RENO,  measure the same $\vert\Delta m^2_{ee}\vert$ for both orderings with uncertainties much larger than JUNO's uncertainty. 

\begin{figure}[b]
  \centering
    \includegraphics[width=0.75\textwidth]{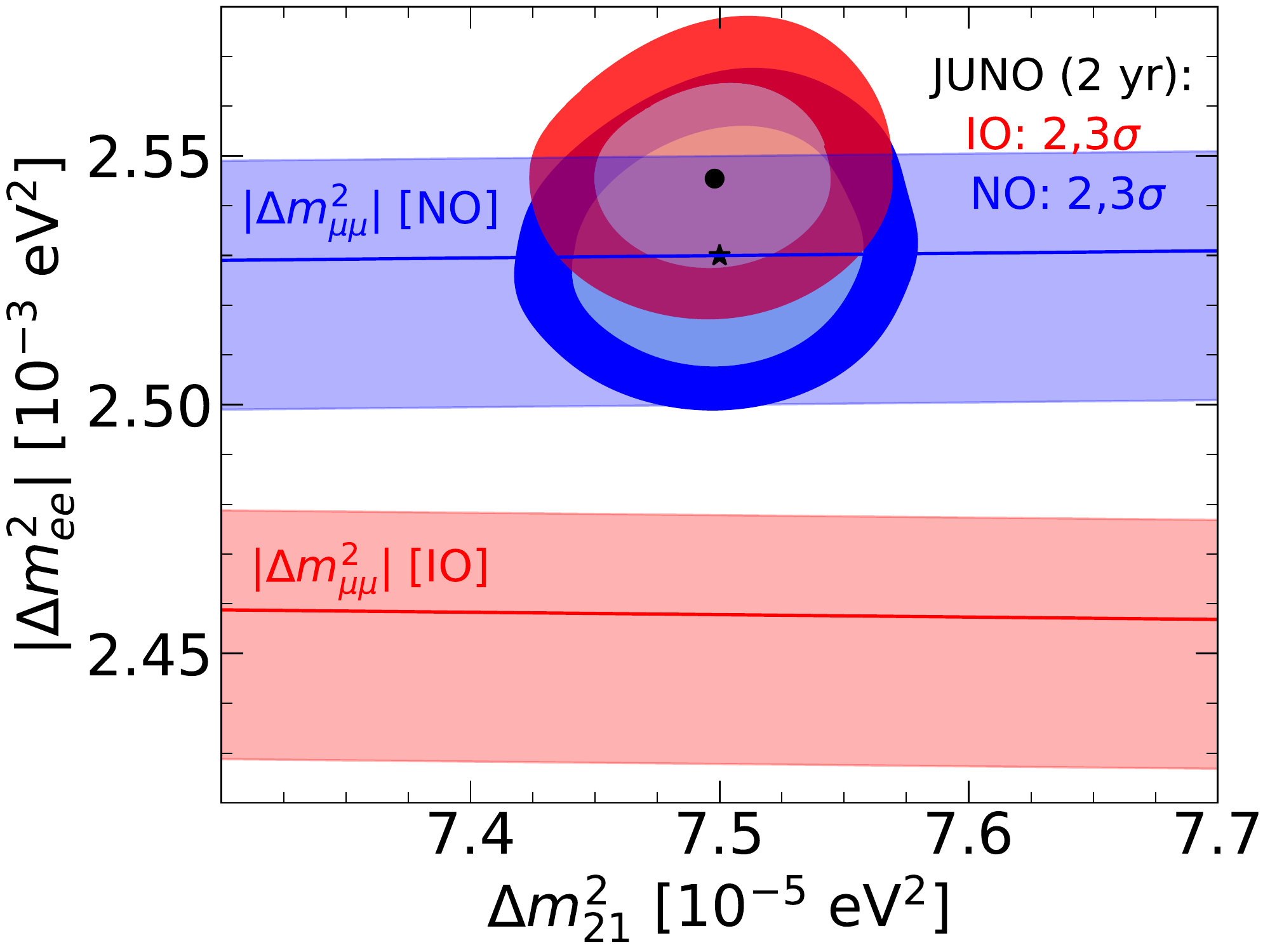}
      \caption{The ellipses are the allowed regions  for JUNO in the $\Delta m^2_{21}$ versus $\vert\Delta m^2_{ee}\vert$ plane for NO (blue, 2 and 3$\sigma$ CL) and IO (red, 2 and 3$\sigma$ CL) after 2 years.  The best fit for NO (IO) is depicted by a black star (dot). We assume NO here and the $\Delta \chi^2$'s are determined with respect to NO best fit point.  We use the neutrino oscillation parameters at the values given in Tab.~\ref{tab:oscparam} and take into account the experimental nominal systematic uncertainties and energy resolution given in Tab.~\ref{tab:sys}. We also show, as red (for IO) and blue (for NO) bands, the 1$\sigma$ CL allowed regions by the current global fit constraint on $\vert \Delta m^2_{\mu \mu}\vert$. Note, the not overlap for the allowed regions for IO.  }
  \label{fig:dm21dmee}
\end{figure}

 This physics is illustrated in Fig.~\ref{fig:dm21dmee} where we show the allowed region in the plane $\Delta m^2_{21}$ versus $\vert\Delta m^2_{ee}\vert$ by  JUNO  for NO (blue) and IO (red) after  2 years of data taking and the corresponding 1$\sigma$ CL  allowed region by the current global fit constraint on $\vert \Delta m^2_{\mu \mu}\vert$. We see that the global fit and JUNO NO regions overlap while the corresponding IO regions do not.
This tension between the position of the best fit values of $\vert \Delta m^2_{ee}\vert$ for IO with respect to NO gives  extra leverage to the data combination.

\begin{figure}[t]
  \centering
  \includegraphics[width=0.45\textwidth]{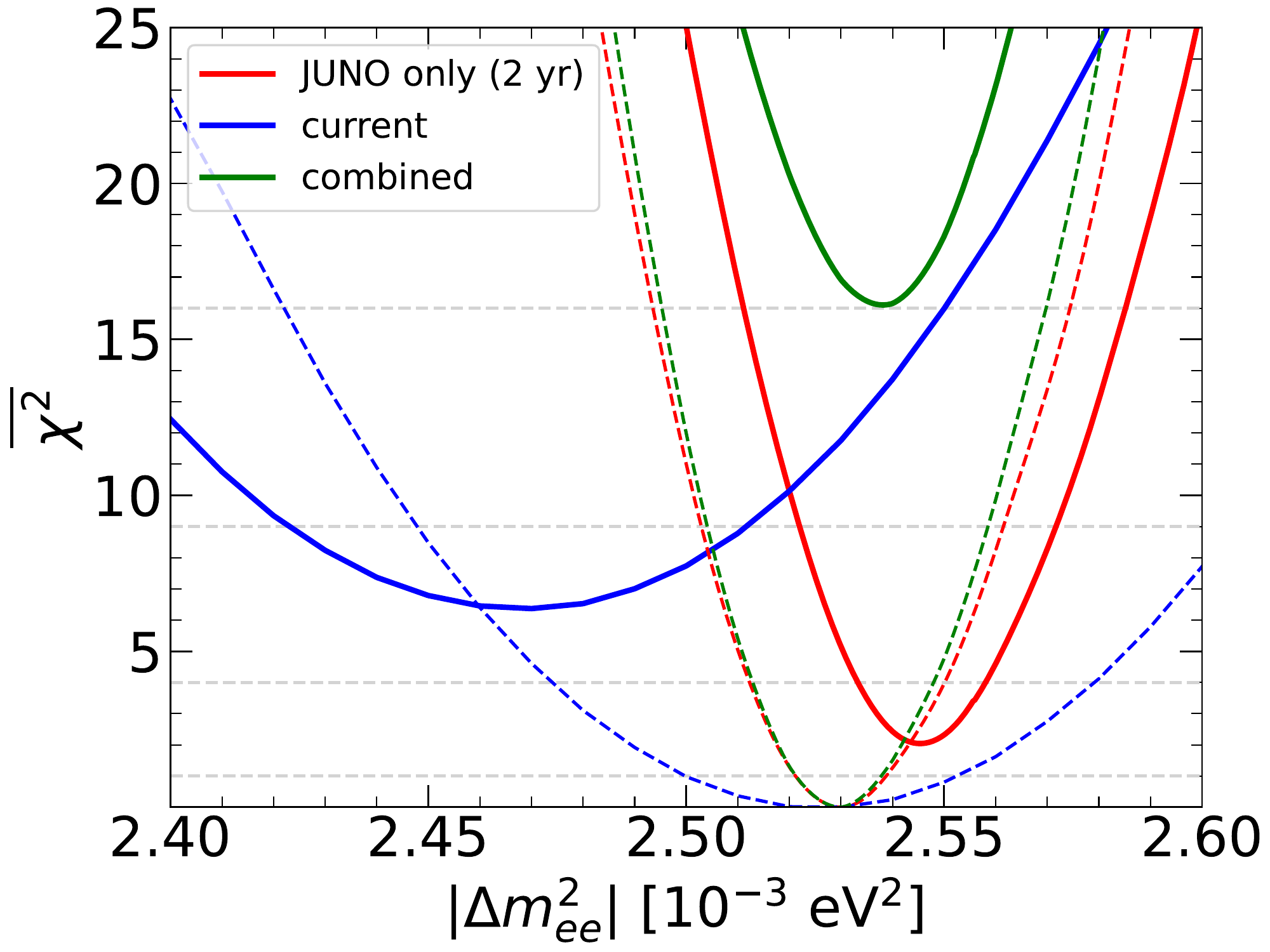}
    \includegraphics[width=0.45\textwidth]{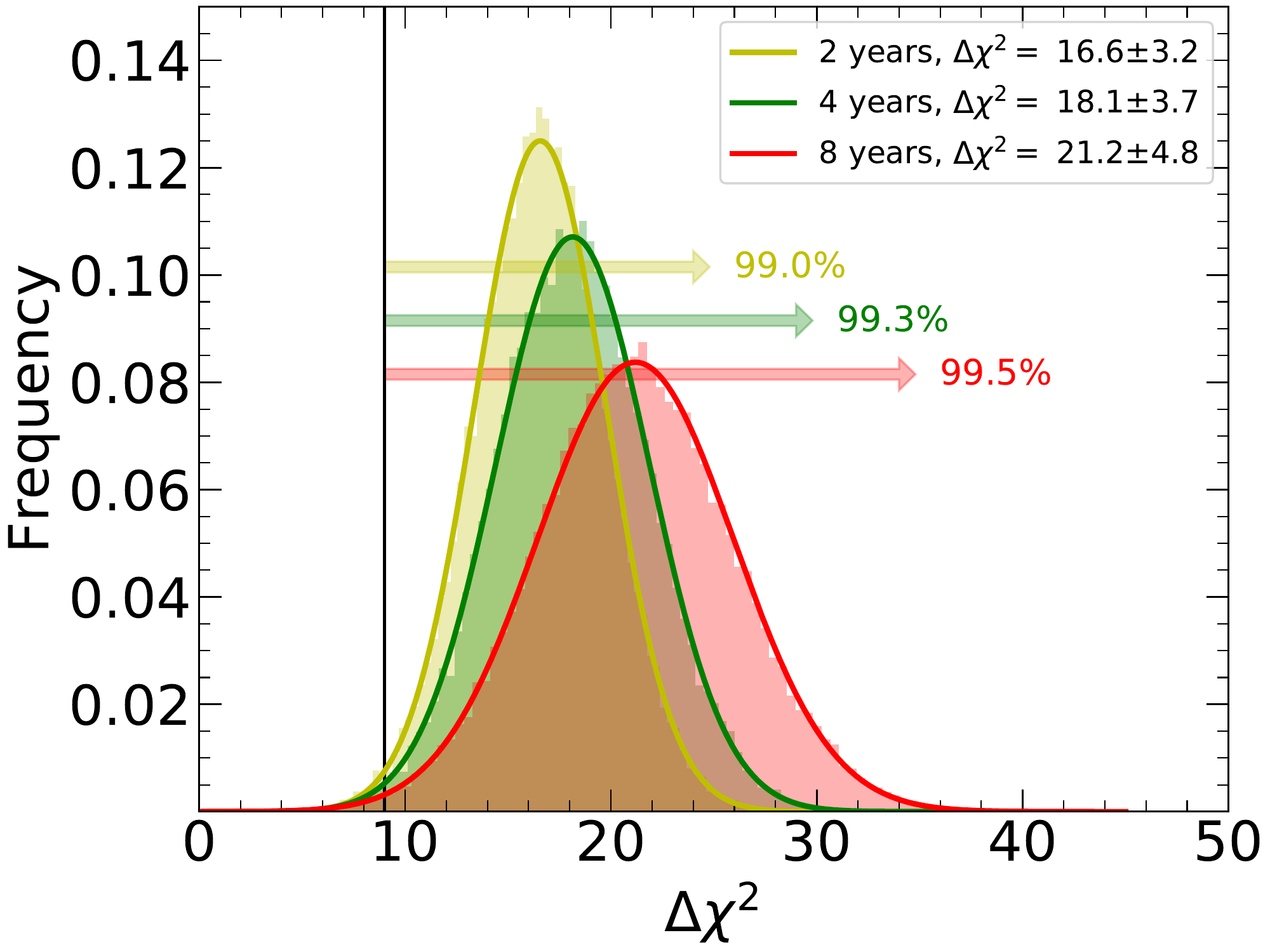}
      \caption{ On the left panel, we show the mean  $\chi^2$ distributions, $\overline{\chi^2}$, for the {\it current} global fit, our predictions for
  JUNO after 2 years.    NO fits shown as dashed lines are the assumed true mass ordering. The fits for IO are shown as solid lines. On the right panel we show 
   the distributions of the $\Delta\chi^2 \equiv \chi^2_\text{min}[\rm IO]- \chi^2_\text{min}[\rm NO]$ values obtained combining the current global fit $\chi^2$ distributions with 
    60\,k trial pseudo-experiments where statistical fluctuations of the trial data have been taken into account for 
    three different exposures: 2 (yellow), 4 (green) and 8 (red) years. For JUNO
    we use the neutrino oscillation parameters at the values give in Tab.~\ref{tab:oscparam} and take into account the experimental nominal systematic uncertainties and energy resolution given in Tab.~\ref{tab:sys}. }
  \label{fig:combined}
 \end{figure}

Therefore, combining JUNO's measurement of $\vert \Delta m^2_{ ee}\vert$ with other experiments, in particular T2K and NOvA, expressed by the current  global fits,  see Refs.~\cite{deSalas:2020pgw,Capozzi:2021fjo,Esteban:2020cvm}, turns out to be very powerful in unraveling the neutrino mass ordering at a high confidence level, as shown in the left panel of Fig.~\ref{fig:combined} for 2 years of JUNO data.
As we can see $\overline{\chi^2}_{\rm min}[\rm IO]$ combined (green solid line) turns out to be about 16.
As a result with only two years of JUNO data taking the mass ordering is determined at better than 3$\sigma$ in 99\% of the trials, see  right panel of  Fig.~\ref{fig:combined}. 
Of course, the actual value of $\overline{\chi^2}_{\rm min}[\rm IO]$ will depend on the value of  $\vert \Delta m^2_{ee}\vert$ measured by JUNO and the updates of the other experiments used in the global fit. In Appendix~\ref{appx:contrib} we discuss the separate contributions  from T2K, NOvA and the atmospheric neutrino data (Super-Kamiokande and DeepCore) to  the $\overline{\chi}^2$ distribution for the global fit determination of $\vert \Delta m^2_{ee}\vert$ and the corresponding impact on the combination with JUNO,  
for completeness and comparison with Fig.~5 of Ref.~\cite{Cabrera:2020own}.

So even though JUNO cannot determine the ordering alone, a couple of years after the start of the experiment, it's precise measurement of  $\vert \Delta m^2_{ ee}\vert$ will allow us to know the mass ordering at better than  3$\sigma$ when the measurement on  $\vert \Delta m^2_{\mu\mu}\vert$ from other neutrino oscillation experiments is combined in a global analysis.

\section{Conclusions}
\label{sec:conc}

The neutrino mass ordering is one of the most pressing open questions in neutrino physics.  It will be most likely measured at different experiments, using atmospheric neutrinos at ORCA~\cite{Adrian-Martinez:2016fdl,Capozzi:2017syc}, PINGU~\cite{Aartsen:2014oha,Winter:2013ema,Capozzi:2015bxa}, Hyper-K~\cite{Abe:2018uyc} or DUNE~\cite{Ternes:2019sak} or accelerator neutrinos at T2HK~\cite{Ishida:2013kba} or DUNE~\cite{Abi:2020qib}. It  also is a flagship measurement for the up-coming JUNO experiment.
This is why we have examined here in detail the impact of various factors on  the determination power of the neutrino mass ordering by JUNO. 

We have assumed NO as the true mass ordering, but our general conclusions do not depend on this assumption. In this case the power of discrimination can be encoded on the value of $\overline{\chi^2}_{\rm min}[\rm IO]$, the larger it is the larger the confidence level one can discriminate between the two mass orderings using JUNO.

We have determined that the real reactor distribution and backgrounds account for a reduction in sensitivity of more than 5 units ({\it i.e.} $\overline{\chi^2}_{\rm min}[\rm IO]$ going from 14.5 to 9.1), the bin to bin flux uncertainty, at its nominal value of 1\%, to an extra reduction of 0.6 down to $\overline{\chi^2}_{\rm min}[\rm IO]=8.5$, both assuming 3\% energy resolution and 200 energy bins.
Note that an improvement on the energy resolution from 3\% to 2.9\%, a  challenging feat to achieve, would represent an increase of $\overline{\chi^2}_{\rm min}[\rm IO]$  from 8.5 to 9.7.

The values of neutrino oscillation parameters that will impact JUNO's measurement are currently known within a few \% uncertainty. 
We have determined the effect of these uncertainties on the mass ordering discrimination. We remark, in particular, the influence of the true value of $\Delta m^2_{21}$, a smaller (larger) value than the current bet fit could shift $\overline{\chi^2}_{\rm min}[\rm IO]$ from 8.5 to 7.1 (10.2).
Another important factor is the non-linear energy response of the detector. Assuming a bias of 0.7\% we have verified that this would decrease $\overline{\chi^2}_{\rm min}[\rm IO]$ further from 8.5 to 7.2.

We have also examined the consequence of statistical fluctuations of the data by performing 60\,k Monte Carlo simulated JUNO pseudo-experiments. Using them we have determined that after 8 (16) years in only 31\% (71\%) of the trials JUNO can determined the neutrino mass ordering at 3$\sigma$ or more. This means that JUNO by itself  will have difficulty determining the mass ordering. 
However, JUNO can still be used for a plethora of different interesting physics analysis~\cite{An:2015jdp,Ohlsson:2013nna,Khan:2013hva,Li:2014rya,Bakhti:2014pva,Chan:2015mca,Abrahao:2015rba,Liao:2017awz,Li:2018jgd,Anamiati:2019maf,Porto-Silva:2020gma,deGouvea:2020hfl,Cheng:2020jje}.
In particular, JUNO will be able to measure $\Delta m^2_{21}$, $\sin^2 \theta_{12}$ and $\vert \Delta m^2_{ee}\vert $ with unmatched precision. 
 This will be very useful to improve our understanding of the pattern of neutrino oscillations and to guide future experiments. 
%

Finally, this inauspicious prediction is mitigated by combining JUNO's $|\Delta m^2_{ee}|$ measurement into the current global fits, in particular the measurement of $|\Delta m^2_{\mu\mu}|$. As we have shown, this combination will most likely result in the determination of the mass ordering at better than 3$\sigma$ with only two years of JUNO data. Our conclusion for the global fits result is consistent with the results of \cite{Cabrera:2020own}.
So we can predict that in approximately two years  after the start of JUNO we will finally know, via global analyses,  the order of the neutrino mass spectrum, i.e. whether the lightest neutrino  mass eigenstate has the most $\nu_e$ ($\nu_1$)  or the least $\nu_e$ ($\nu_3$).   \\

\begin{acknowledgements}
We would like to thank Pedro Machado 
for useful comments on a preliminary version of this paper.
CAT and RZF are very thankful for the hospitality of the Fermilab Theoretical Physics Department, where this work was initiated. 
Fermilab is operated by the Fermi Research Alliance under contract no.~DE-AC02-07CH11359 with the U.S.~Department of Energy.
CAT is supported by the research grant
``The Dark Universe: A Synergic Multimessenger Approach'' number 2017X7X85K under the program ``PRIN 2017'' funded by the Ministero dell'Istruzione, Universit\`a e della Ricerca (MIUR). 
RZF is partially supported by Funda\c{c}\~ao de Amparo \`a Pesquisa do Estado de S\~ao Paulo (FAPESP) and Conselho Nacional de Ci\^encia e Tecnologia (CNPq).
This project has received funding/support from the European Union's Horizon 2020 research and innovation programme under the Marie Sklodowska-Curie grant agreement No 860881-HIDDeN.

\end{acknowledgements}

\appendix

\newpage

\section{Artificial Constraints}
\label{appx:artificial}

Using  $ \Delta m^2_{ee} ~[\text{NO}] = 2.530  \times 10^{-3} $ eV$^2$:\\

\begin{enumerate}
\item then for the artificial constraint that   $|\Delta m^2_{32} ~[\text{IO}]|    = \Delta m^2_{32} ~[\text{NO}] $   \cite{Petcov:2001sy}
$$ |\Delta m^2_{ee} ~[\text{IO}]|    = \Delta m^2_{ee} ~[\text{NO}]  { -  2 \cos^2 \theta_{12} \Delta m^2_{21} = 2.428 \times 10^{-3} ~ \text{eV}^2  }$$  

\item  then for the artificial constraint that    $|\Delta m^2_{31} ~[\text{IO}]|    = \Delta m^2_{31} ~[\text{NO}] $   \cite{Choubey:2003qx}
$$|\Delta m^2_{ee} ~[\text{IO}]|    = \Delta m^2_{ee} ~[\text{NO}]  + { 2 \sin^2 \theta_{12} \Delta m^2_{21}} = 2.578 \times 10^{-3} ~ \text{eV}^2  $$

\item   then for the artificial constraint that   $ |\Delta  m^2_{32} ~[\text{IO}]|    = \Delta m^2_{31} ~[\text{NO}] $  
 \cite{Bilenky:2017rzu}  
 (See also Ref.~  \cite{Tanabashi:2018oca}, 
 Neutrino review,  Section 14.7, eq. 14.48.).
 $$|\Delta m^2_{ee} ~[\text{IO}]|    = \Delta m^2_{ee} ~[\text{NO}]  - \cos2\theta_{12} \Delta m^2_{21} = 2.503 \times 10^{-3}~  \text{eV}^2 $$

\end{enumerate}

The actual $\chi^2$ minimum, obtained numerically in  Fig. \ref{fig:spectra}, is when
$$|\Delta m^2_{ee} ~[\text{IO}]|    \approx  2.548 \times 10^{-3}  ~\text{eV}^2  $$		
i.e.  midway between   the $|\Delta m^2_{31} ~[\text{IO}]|    = \Delta m^2_{31} ~[\text{NO}] $  and the  $|\Delta  m^2_{ee} ~[\text{IO}]|    = \Delta m^2_{ee} ~[\text{NO}] $ artificial constraints.
It is also easy to see from  Fig. \ref{fig:spectra} 
that imposing any of these artificial constraints {\it significantly increases} the size of the $\overline{\Delta \chi^2}$ between the fits of the two mass orderings and therefore gives misleading confidence levels for the determination of the neutrino mass ordering. Note that all of the below give equivalent $\Delta m^2_{ij}$'s :
\begin{align}
\Delta m^2_{ee} ~[\text{NO}] = 2.530  \times 10^{-3} ~ \text{eV}^2, \quad  &  |\Delta m^2_{ee} ~[\text{IO}]|    \approx  2.548 \times 10^{-3}  ~\text{eV}^2,  \\
\Delta m^2_{32} ~[\text{NO}] = 2.479  \times 10^{-3} ~ \text{eV}^2, \quad   &  |\Delta m^2_{32} ~[\text{IO}]|    \approx  2.581 \times 10^{-3}  ~\text{eV}^2,  \\
\Delta m^2_{31} ~[\text{NO}] = 2.554  \times 10^{-3} ~  \text{eV}^2, \quad   &  |\Delta m^2_{31} ~[\text{IO}]|    \approx  2.506 \times 10^{-3}  ~\text{eV}^2\,.  
\end{align}
When minimizing the $\chi^2$ difference for Fig. \ref{fig:spectra},  the change in ($|\Delta m^2_{ee}|$, $|\Delta m^2_{32}|$, $|\Delta m^2_{31}|$) going from NO to IO is (+0.7\%, +4.0\%, -1.9\%) respectively, i.e. the minimal difference is for $|\Delta m^2_{ee}|$.

\newpage

\section{$\nu_e$ Disappearance Probability in Vacuum}
\label{appx:Prob}

\begin{figure}[t]
  \centering
  \includegraphics[width=0.45\textwidth]{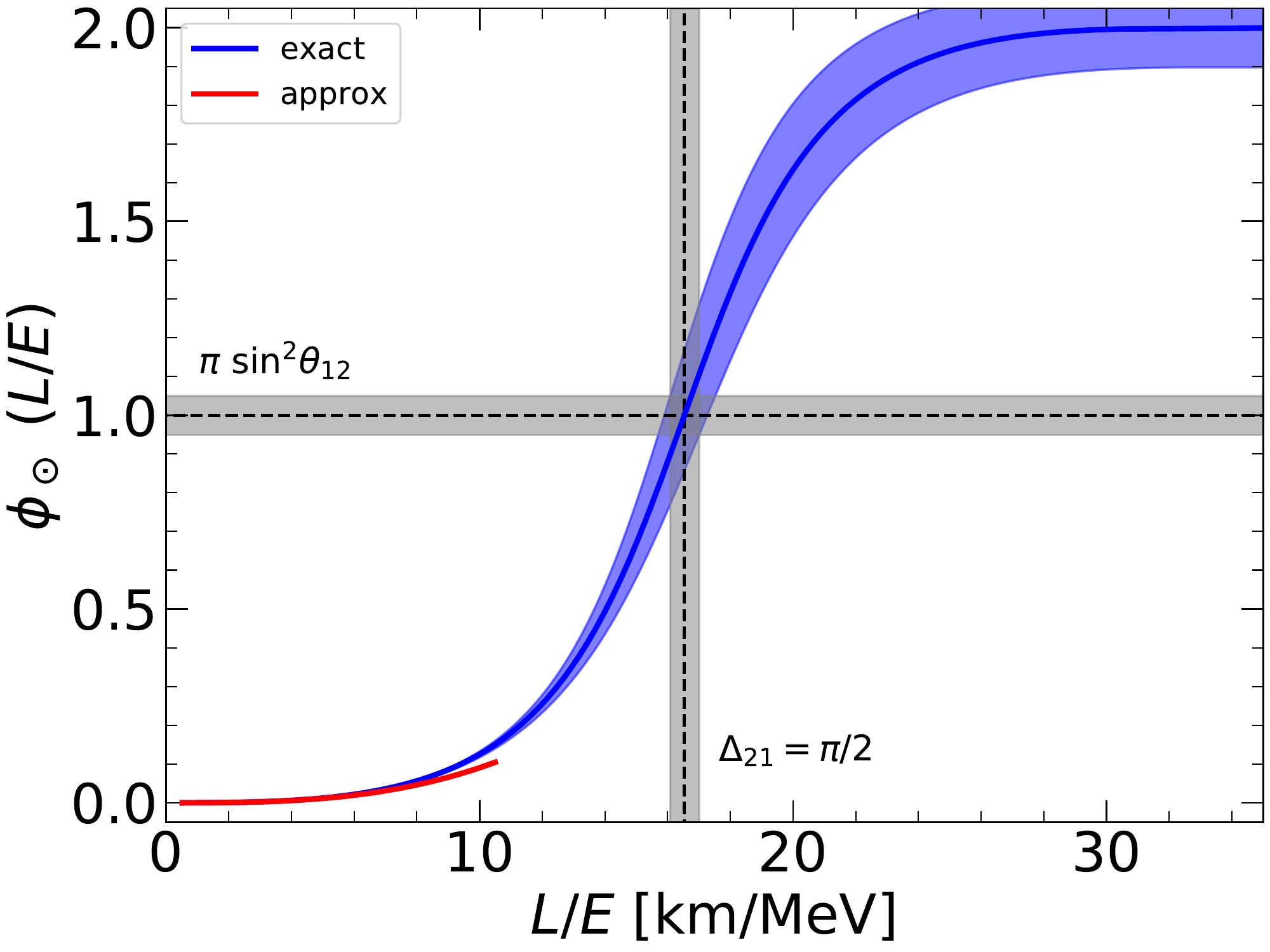}
    \caption{The kinematic phase advance/retardation for the survival probability,  $\Phi_\odot$, as a function of $L/E$ (left) and $E$  at $L=52.5$~km (right). 
    The blue band is obtained from the exact formula, while the red curve shows the approximation for  values of $L/E< 10$ km/MeV. 
The dashed vertical and horizontal lines mark the  solar oscillation minimum, i.e. $\Delta _{21} = \pi/2$ where $\Phi_\odot = \pi~\sin^2\theta_{12}\approx 0.999$. The gray bands are obtained by varying the solar parameters in their corresponding 1$\sigma$ intervals as given in Table \ref{tab:oscparam}. }
  \label{fig:phi_figb}
\end{figure}

We start from the usual expression for the $\nu_e$ disappearance probability in vacuum, 
\begin{eqnarray}
P_{\overline{\nu}_e\to\overline{\nu}_e} = 1 &-&\sin^22\theta_{12}\cos^4\theta_{13}\sin^2\Delta_{21}\nonumber
 \\&-&
 \sin^22\theta_{13}\left[\cos^2\theta_{12} \sin^2 \Delta_{31} +\sin^2 \theta_{12} \sin^2 \Delta_{32} \right] \, .
 \label{eq:prob0_app}
\end{eqnarray}
Using the methods from Ref.~\cite{Parke:2016joa}, the simplest way to show that
\begin{eqnarray}
\cos^2\theta_{12} \sin^2 \Delta_{31}+\sin^2\theta_{12} \sin^2 \Delta_{32} & = & \frac{1}{2}  \biggl( 1 - \sqrt{ 1- \sin^2 2 \theta_{12} \sin^2 \Delta_{21}  } ~ \cos \Omega \biggr) 
\end{eqnarray}
 with 
\begin{eqnarray} 
\Omega & = & 2 \Delta_{ee} + {\Phi_\odot },  \\[3mm]
{\rm where}  \quad  \Delta m^2_{ee}  & \equiv  & \frac{\partial ~\Omega}{\partial (L/2E)} \left|_{\frac{L}{E} \rightarrow 0} \right.
 =  \cos^2 \theta_{12}\Delta m^2_{31}+\sin^2 \theta_{12} \Delta m^2_{32} 
\\[3mm]
{\rm and}  \quad \quad { \Phi_\odot } & \equiv  & \Omega - 2 \Delta_{ee} 
= \arctan(\cos 2 \theta_{12} \tan \Delta_{21}) - \Delta_{21} \cos 2 \theta_{12},
\end{eqnarray}
as shown in Fig. \ref{fig:phi_figb}, is to write  
\begin{eqnarray}
c^2_{12} \sin^2 \Delta_{31}+s^2_{12} \sin^2 \Delta_{32} &= & \frac{1}{2}  \biggl( 1- (c^2_{12} \cos 2\Delta_{31}+s^2_{12} \cos 2\Delta_{32}) \biggr),
\end{eqnarray}
using $c^2_{12} \equiv \cos^2 \theta_{12}$ and $s^2_{12} \equiv \sin^2 \theta_{12}$. 
Then, if we rewrite $2\Delta_{31}$ and $2\Delta_{32}$ in terms of $(\Delta_{31}+\Delta_{32})$ and $\Delta_{21}$, 
we have
\begin{eqnarray}
c^2_{12} \cos 2\Delta_{31}+s^2_{12} \cos 2\Delta_{32} & = & 
c^2_{12} \cos (\Delta_{31}+\Delta_{32}+\Delta_{21})+s^2_{12} \cos (\Delta_{31}+\Delta_{32}-\Delta_{21}) \nonumber \\
& = & \cos(\Delta_{31}+\Delta_{32}) \cos \Delta_{21} - \sin(\Delta_{31}+\Delta_{32}) \cos 2 \theta_{12} \sin \Delta_{21}.  \nonumber 
\end{eqnarray}
Since
\begin{eqnarray}
 \cos^2 \Delta_{21}+ \cos^2 2 \theta_{12} \sin^2 \Delta_{21} =1- \sin^2 2 \theta_{12} \sin^2 \Delta_{21} \nonumber
 \end{eqnarray}
we can then write
\begin{eqnarray}
c^2_{12} \cos 2\Delta_{31}+s^2_{12} \cos 2\Delta_{32} 
& = & \sqrt{ 1- \sin^2 2 \theta_{12} \sin^2 \Delta_{21}  }  ~ \cos \Omega,   
\label{eq:omega}
\end{eqnarray}
where
\begin{eqnarray}
\Omega & = & \Delta_{31}+ \Delta_{32} + \arctan(\cos2\theta_{12} \tan \Delta_{21}). \nonumber
\end{eqnarray}
To separate $\Omega$ into an effective $2\Delta$ and a phase, $\Phi_\odot$, we have
\begin{eqnarray}
 \frac{\partial ~\Omega}{\partial (L/2E)} \left|_{\frac{L}{E} \rightarrow 0} \right.
& =  &   \cos^2 \theta_{12}\Delta m^2_{31}+\sin^2 \theta_{12} \Delta m^2_{32} =\Delta m^2_{ee}  \nonumber \\[3mm]
{\rm and} \quad \Phi_\odot  & = &  \Omega - 2 \Delta_{ee}  
=  \arctan(\cos 2 \theta_{12} \tan \Delta_{21}) - \Delta_{21} \cos 2 \theta_{12}\,.   \nonumber
\end{eqnarray}
Thus
\begin{eqnarray}
\Omega & = & 2 \Delta_{ee} +  (\arctan(\cos2\theta_{12} \tan \Delta_{21})  -  \Delta_{21} \cos2\theta_{12}).
\end{eqnarray}
Since $\Omega$  appears only as $\cos \Omega$, one could use  $\Omega= 2 |\Delta_{ee}| \pm \Phi_\odot$ as in Eq.~\eqref{eq:prob}.

The factor  $\sqrt{ 1- \sin^2 2 \theta_{12} \sin^2 \Delta_{21}  }$ in front of $\cos \Omega $ in Eq. \eqref{eq:omega}, modulates the amplitude of the $\theta_{13}$ oscillations 
as this factor varies from 1 to $\cos 2 \theta_{12} \approx 0.4$ as $\Delta_{21}$ goes from 0 to $\pi/2$.
So the $\sqrt{(\cdots)}$ modulates the amplitude and $\Phi_\odot$ modulates the phase of the $\theta_{13}$ oscillations.


\section{Verification of our code}
\label{app:compare}

In this appendix, we show that using our code we can reproduce former results obtained by the JUNO collaboration. 
In particular, we compare with the results from Ref.~\cite{Bezerra:2019dao}.
Note that some of the experimental features have improved since this analysis has been performed, in particular the overall detection efficiency and a reduction of accidental background events.

\begin{figure}[t]
  \centering
  \includegraphics[width=0.43\textwidth]{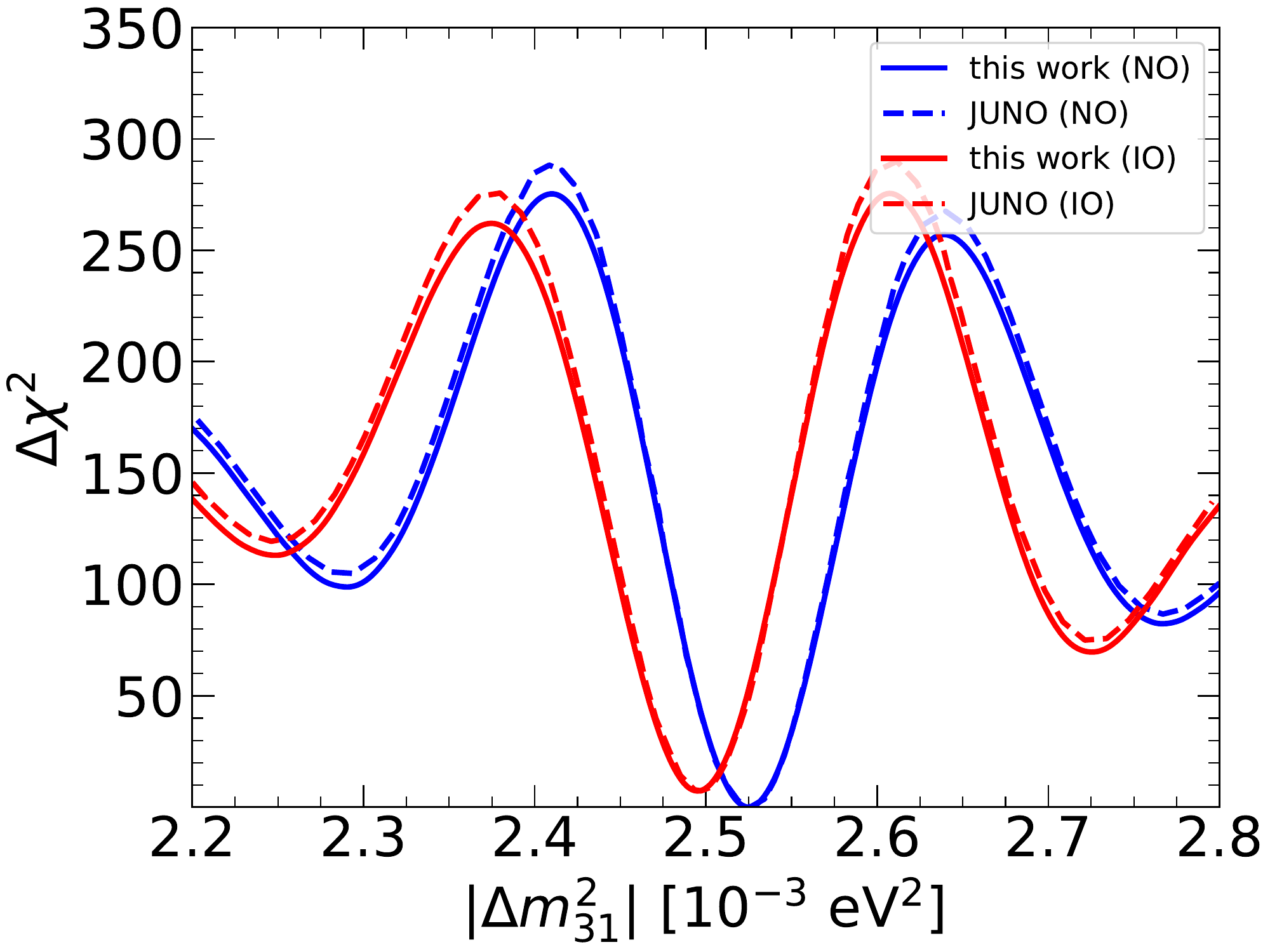}
  \includegraphics[width=0.43\textwidth]{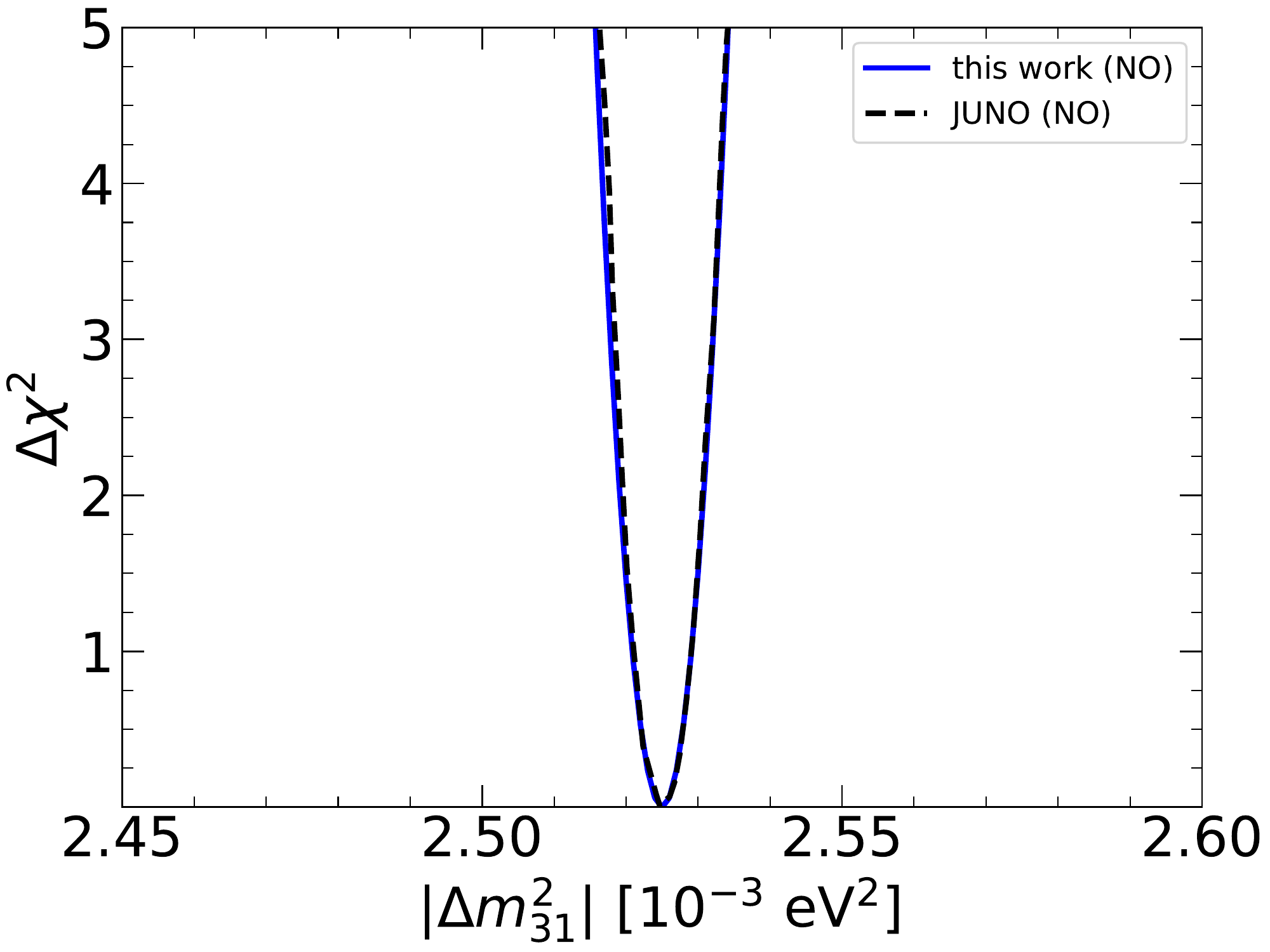}
    \caption{ Here we reproduce Figs. 4 and 11 from  Ref.~\cite{Bezerra:2019dao}, using the oscillation parameters and technical details of that reference.
  Our code, written for this paper, gives the solid lines whereas the results extracted from the above reference are dashed lines, normal (inverted) ordering is in blue (red).}
  \label{fig:compare}
\end{figure}

We assume 6 years of exposure time (1800 days). No NL effects are included in the analysis and the 1\% shape uncertainty is included as a modification of the denominator of the $\chi^2$ function~\cite{Bezerra:2019dao}. In particular, we use for this cross check

\begin{equation}
 \chi^2(\vec{p}) =\min_{\vec{\alpha}}\sum_{i} \frac{(N_{i}^\text{dat} - N_{i}(\vec{p},\vec{\alpha}))^2}{N_{i}(\vec{p},\vec{\alpha}) + \sigma_s^2 N_{i}(\vec{p},\vec{\alpha})^2} + \sum_j\left(\frac{\alpha_j}{\sigma_j}\right)^2,
\end{equation}
in accordance with Ref.~\cite{Bezerra:2019dao}, but slightly different to our Eq.~\eqref{eq:chi2}. Here, $\sigma_s = 0.01$.
In Fig.~\ref{fig:compare} we compare the results from our analysis (dashed lines) with the lines extracted directly from Ref.~\cite{Bezerra:2019dao} (solid lines).
As can be seen the results agree very well with each other. In perfect agreement with the collaboration, we obtain $\chi^2_{\rm min}[{\text{IO}}] = 7.3$.

%
%
%

\section{On the contribution to the determination of $\vert\Delta m^2_{ee}\vert$ from the $\vert\Delta m^2_{\mu \mu}\vert$ sensitive experiments}
\label{appx:contrib}

\begin{figure}[h]
  \centering
  \includegraphics[width=0.49\textwidth]{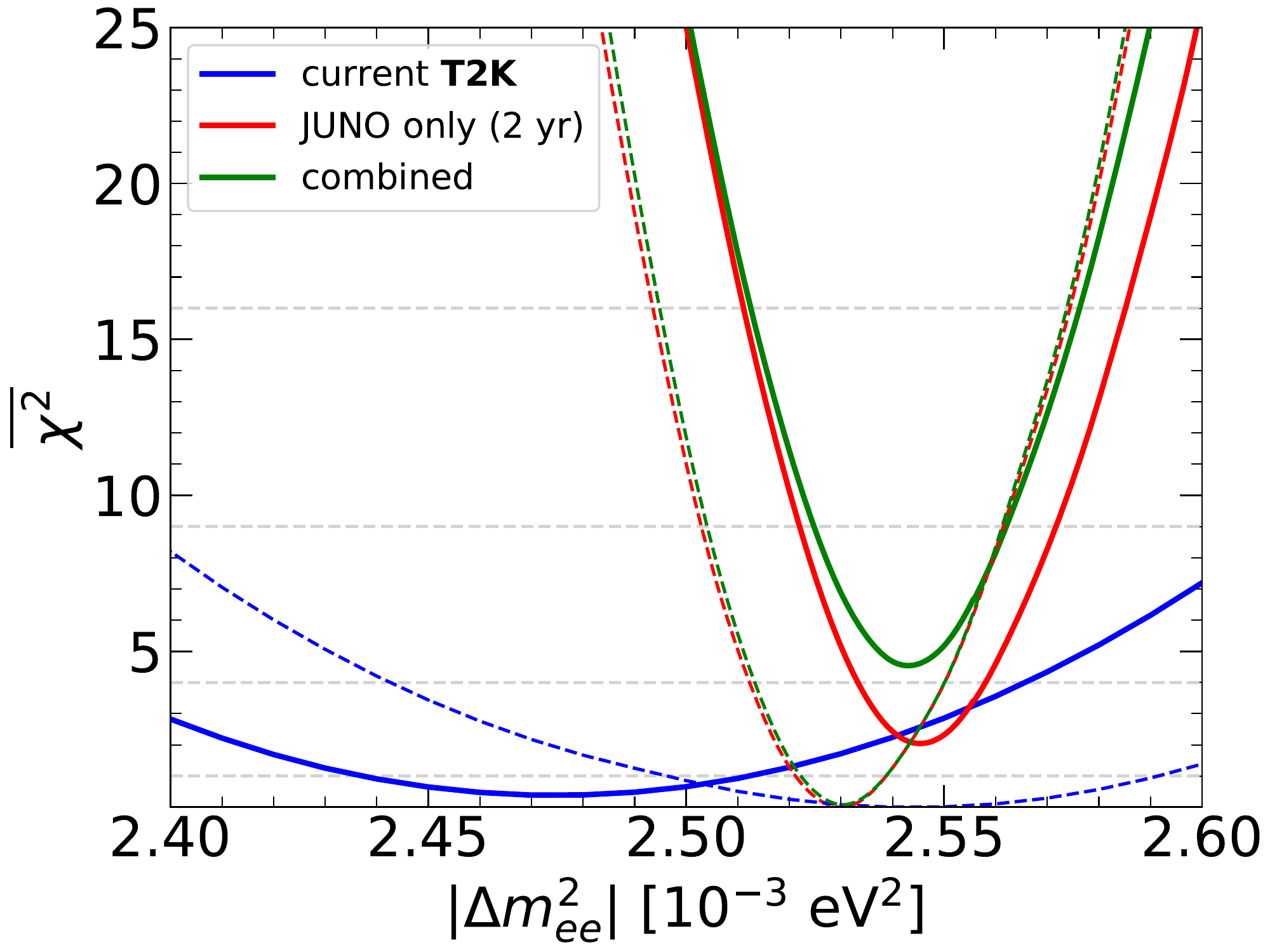}
  \includegraphics[width=0.49\textwidth]{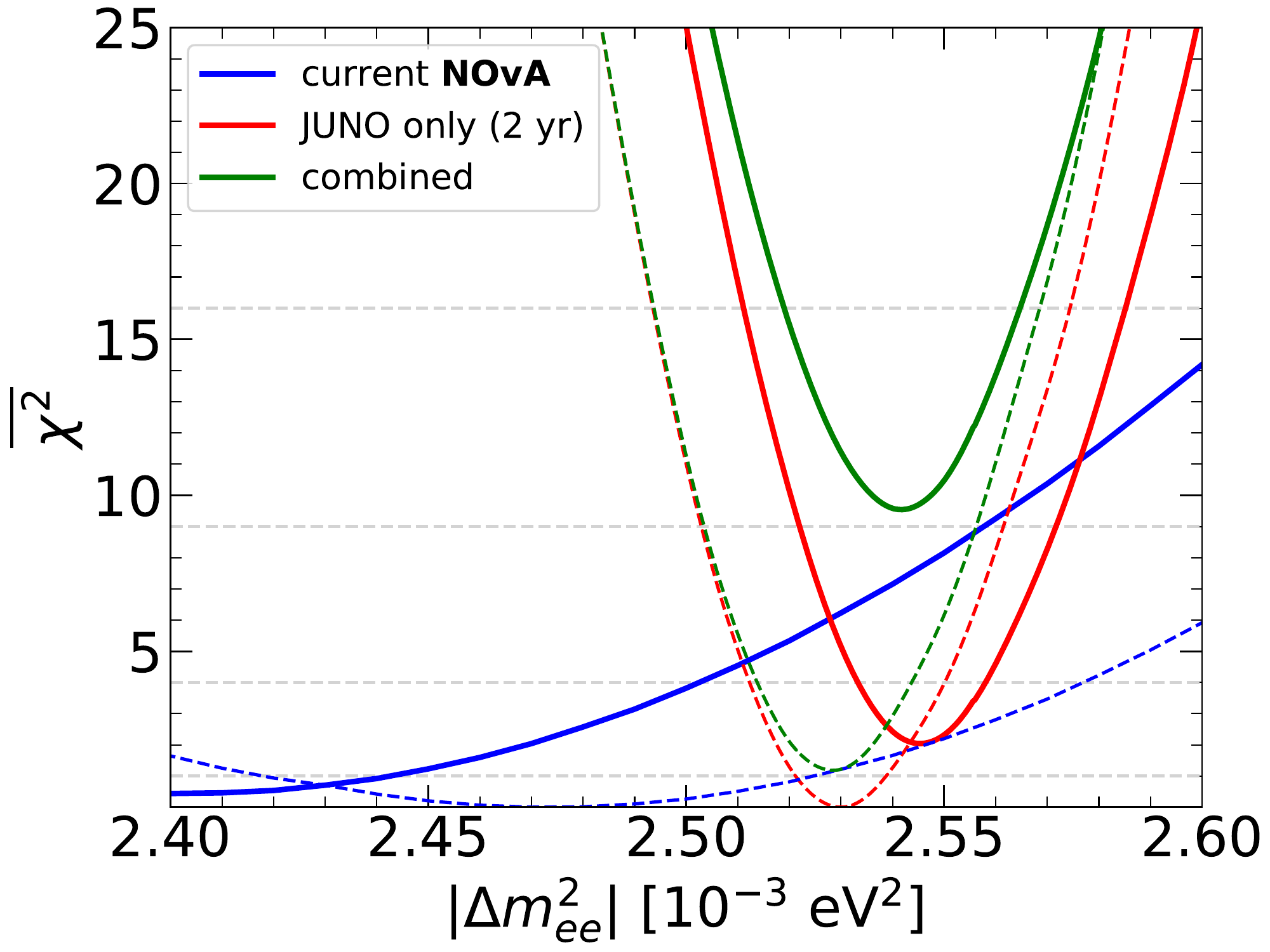} \\[0.5cm]
   \includegraphics[width=0.49\textwidth]{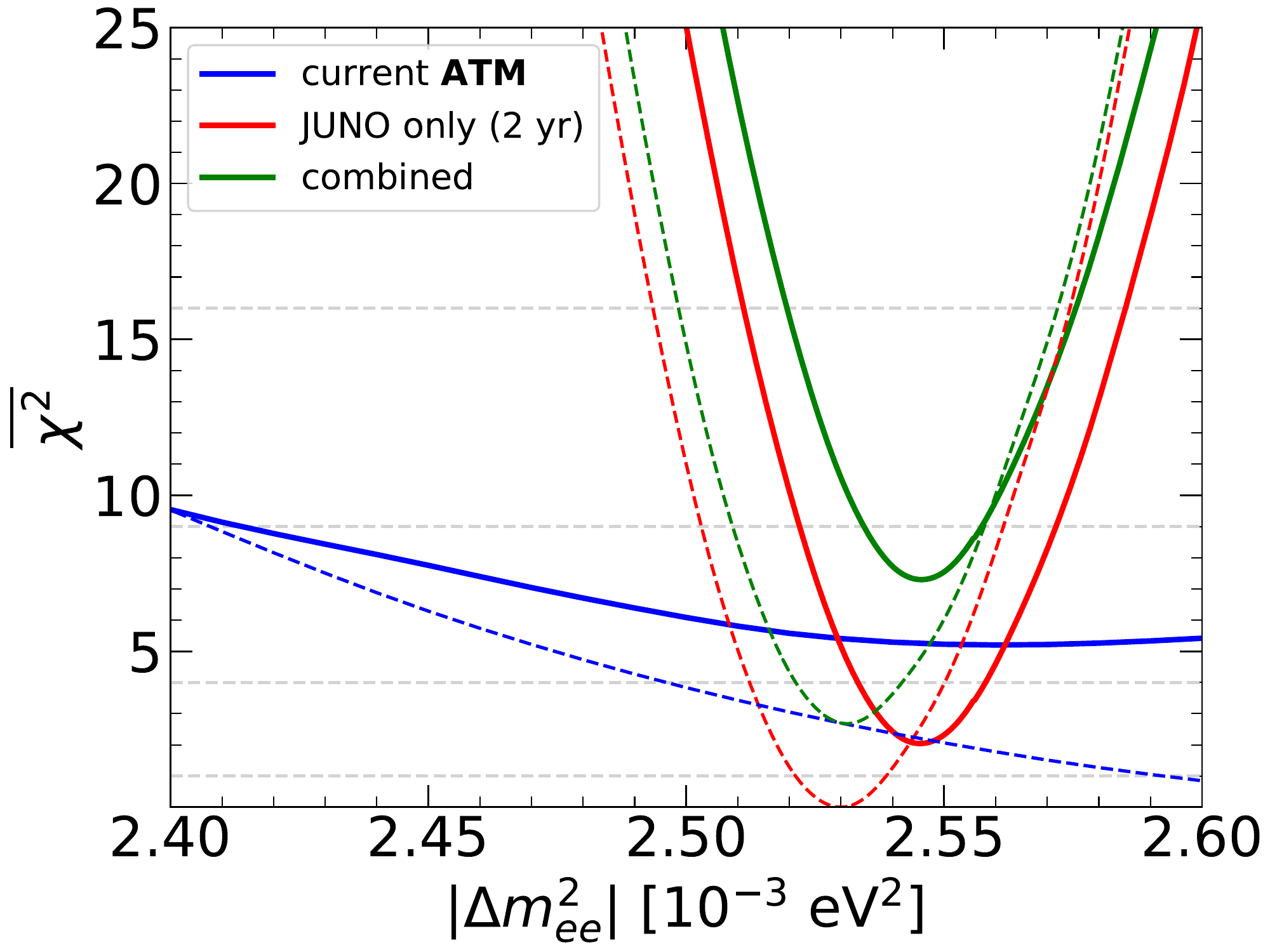}
    \caption{Separate contributions of T2K data (upper left panel), NOvA data (upper right panel) and Super-K and DeepCore atmospheric data, labeled  ATM, (lower panel) to the $\overline \chi^2$  fit of 
    $\vert \Delta m^2_{ee}\vert$  to NO (dashed lines)  and IO (solid lines) included in the global fit (blue) and in the combination of the current global fit with  2 years of JUNO data (green). JUNO fit only is in red.}
  \label{fig:t2knova}
\end{figure}

 It is informative to examine the contributions of the $\vert\Delta m^2_{\mu \mu}\vert$ sensitive experiments included in the global fit to the final determination of 
$\vert\Delta m^2_{ee}\vert$. We will focus here on the major players: T2K, NOvA and the atmospheric neutrino oscillation experiments Super-Kamiokande and DeepCore (ATM).  The analyses of T2K, NOvA and ATM data shown in this section correspond to the analyses performed in Ref.~\cite{deSalas:2020pgw}.
For this purpose we show in Fig.~\ref{fig:t2knova}  the separate contributions to  the determination of $\vert \Delta m^2_{ee}[\rm NO]\vert$ and $\vert \Delta m^2_{ee}[\rm IO]\vert$ coming from  T2K (upper left panel), NOvA (upper right panel) and the ATM  (lower panel) neutrino oscillation data. We show their effect on the global fit and on the corresponding global fit combination with 2 years of JUNO data.

From these plots we see that  T2K prefers $\vert \Delta m^2_{ee}[\rm NO,IO]\vert$  closer to the global fit best fit values, while NOvA (ATM) prefers  lower (higher) values. Note that both accelerator neutrino oscillation experiments, however, prefer $\vert \Delta m^2_{ee}[\rm IO]\vert$ smaller  than the value JUNO will prefer (NO assumed true).
Since none of the $\overline{\chi}^2$ distributions are very Gaussian  at this point, the combined  $\overline{\chi}^2_{\rm min}[\rm IO]$ is a result of  broad distributions pulling for different minima that at JUNO's best fit value for $\vert \Delta m^2_{ee}[\rm IO]\vert$ contribute to an increase of  $\overline{\chi}^2_{\rm min}[\rm IO]$ of about 
7 (NOvA), 3 (T2K) and 5 (ATM) units, resulting on the final power of the combination. 

The addition of the atmospheric data, and also to a minor extent of MINOS data (which is compatible with NOvA), 
to the global fit used in this paper explains the difference of about 4 units in the predicted boost for the determination of the mass ordering we show here with respect to what is predicted in Fig.~5 of Ref.~\cite{Cabrera:2020own}, where only simulated data from T2K and NOvA were used.

\newpage

\bibliographystyle{utphys}
\providecommand{\href}[2]{#2}\begingroup\raggedright\endgroup
\end{document}